\documentclass[aps,pra,reprint,twocolumn,floatfix,groupedaddress,superscriptaddress]{revtex4-2}
\usepackage{pstricks,graphicx,amsmath,bbm,mathrsfs,amssymb,psfrag,pifont,times,mathptmx}
\usepackage[utf8]{inputenc}
\usepackage{hyperref}
\usepackage[english]{babel}
\usepackage{color}
\usepackage{ulem}
\usepackage{siunitx}
\usepackage{mathtools}
\usepackage{comment} % to exclude whole sections when I want to

\def\dag{{^{\dagger}}}

\def\cat{{\psi_{\rm cat}}}
\def\UG{{\rm GP}}
\def\IMP{{\rm imp}}

\def\DISP{{\rm D}}

\def\PD{{\rm pd}}

\begin{document}
\title{Deterministic feedforward-based generation of large optical %Schr\"odinger cat states
 coherent-state superposition}
\author{Michele N.~Notarnicola}
\email{michelenicola.notarnicola@upol.cz}
\affiliation{Department of Optics, Palack\'y University,
17. Listopadu 12, 779 00 Olomouc (Czech Republic) }

\author{Marcin Jarzyna}
\email{m.jarzyna@cent.uw.edu.pl}
\affiliation{Centre for Quantum Optical Technologies, 
Centre of New Technologies, 
University of Warsaw, Banacha 2c, 02-097 Warszawa (Poland)}

\author{Radim Filip}
\email{filip@optics.upol.cz}
\affiliation{Department of Optics, Palack\'y University,
17. Listopadu 12, 779 00 Olomouc (Czech Republic) }

\date{\today}
%%%%%%%%%%%%%%
\begin{abstract}
Large %Schr\"odinger cat states of light 
optical coherent-state superpositions are essential to advance quantum sensing, quantum repeaters and error-correction codes. We propose a deterministic feedforward protocol employing qubit-mode dispersive coupling, currently available in cavity quantum electrodynamics (QED). We show this single-mode protocol to outperform the advanced three-mode Gaussian-photon-number-resolving detector scheme both in terms of average fidelity and quantum non-Gaussian phase-space properties, and propose sensitivity to weak displacements of interference fringes as a feasible and conclusive witness of quantum interference. This approach combining QED with electro-optical feedforward is extendable to tailored states for applications and other platforms.
\end{abstract}
\maketitle
%%%

%%%%%%%%%%%%%%%%%%%%%%%%%%%%%%%%%%%%%%%%%%%%%%%%%%%%%%%%%
%%%%%%%%%%%%%%%%%%%%%%%%%%%%%%%%%%%%%%%%%%%%%%%%%%%%%%%%%
\section{Introduction}

Generation %of Schr\"odinger cat states %on different platforms
of quantum superposition states 
is a challenging task in quantum information science %, not only from a fundamental perspective
%\cite{Arndt2014, Nori2021}, but also for practical purposes, 
as they provide resource for quantum metrology \cite{Joo2011, Huang2015, Tatsuta2019}, %spectroscopy \cite{Kira2011, Milne2021}, 
computation \cite{Ralph2003, Jeong1, Jeong2, Cao2024, Walls2025}, communication \cite{Rep1,Rep2,Rep3,Rep4,Rep5} and continuous-variable quantum error correction \cite{Schlegel2022, Hastrup2022}. 
A special challenge rises for optical coherent-state superposition, hereafter referred to as {\it optical cat states}, %corresponding to superposition of coherent states of radiation with opposite phases, 
namely $|\cat ^{(\pm)}\rangle= (|\alpha\rangle \pm |-\alpha\rangle)/{\cal N}_\pm$, 
%\begin{align}
%|\cat ^{(\pm)}\rangle= \frac{1}{{\cal N}_\pm } \left(|\alpha\rangle \pm |-\alpha\rangle \right) \, , 
%\end{align}
$\alpha\ge0$, with ${\cal N}_\pm= \sqrt{2(1\pm \exp(-2\alpha^2))}$ \cite{Gerry1997}. 
%Even though the seminal ideas for their generation were measurement-free strategies employing self Kerr interaction in $\chi^{(3)}$ nonlinear media \cite{Yurke1986, Yurke1988, Paris1999}, the currently most popular 
Conventional methods adopt probabilistic schemes based on either photon subtraction \cite{Kitten, Takahashi2008, Laghaout2013, Furusawa2021_Generalized} or addition \cite{Chen2024}, that both exploit two- or three-mode Gaussian coupling between signal and ancillas followed by photon-number-resolving detectors (PNRDs), but, despite feasibility, are associated with two main limitations. Firstly, they are efficient only for kitten generation, as they implement conditional projection over a finite superposition of Fock states embedded in a Gaussian envelope \cite{Menzies2009}. In turn, generation of cats with $\alpha\ge 2$ becomes challenging, requiring a large number of heralded photons to achieve high fidelity and, moreover, quantum non-Gaussian (QNG) phase space properties, thus inducing a significant drop of the success rate \cite{Furusawa2021_Generalized}. Even if cat enlargement protocols have also been proposed, they come at the cost of further reduction of success probability \cite{Sychev2017}, that can be partially compensated by high quality quantum memories, whose realization is, however, currently challenging too. Secondly, QNG interference properties of the resulting states are only partially reproducing the desired ones, as they only get finite stellar rank \cite{Biswas2007, Luo2024}. 

\begin{figure}[h!]
\includegraphics[width=0.99\columnwidth]{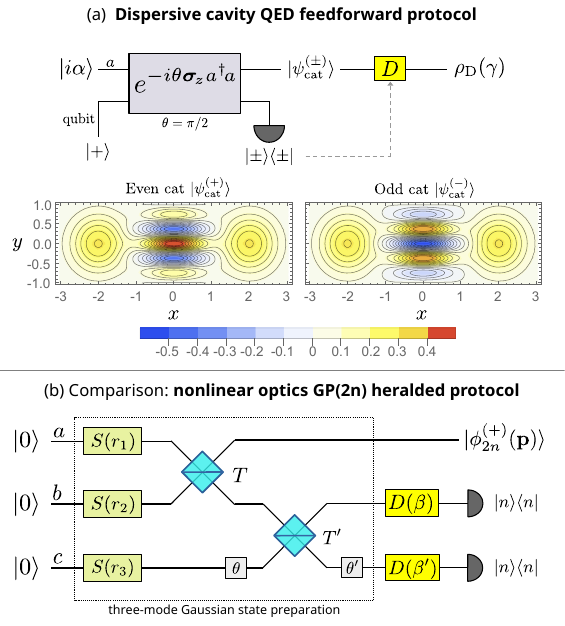}
\caption{(a) Dispersive coupling protocol for even cat generation, based on qubit-light interaction and feedforward displacement conditioned on qubit measurement. The qubit measurement conditionally projects light to either the even or the odd cat state, whose Wigner functions $W_{\rm cat}^{(\pm)}(x,y)$ are plotted in the panels above for $\alpha=2$. If outcome ``-" is retrieved, state $|\cat^{(-)}\rangle$ is displaced along $y$ direction of phase space to make its quantum interference fringes overlap with those of the even cat, thus making the protocol deterministic.
(b) Gaussian-PNRD-based [$\UG(2n)$] protocol, performing conditional displacement-PNRD over an optimized three-mode Gaussian state. Extensions of this basic scheme, including displacements in state preparation and feedforward PNRD, are not helpful to improve quality of the output state \cite{SuppMat}.
%to include  feedforward PNRD is discussed in \cite{SuppMat}, but limits itself to increase success rate without affecting the quality of the output state
}\label{fig:01-EvenGaussianSchemes}
\end{figure}

On the other hand, recent experimental progresses in optical cavity quantum electrodynamics (QED) have proved itself as candidate for the task \cite{Magro2023}, with the advantage of deterministic generation and flexibility to implement different types of atom-light interaction \cite{Wang2005, Welte2018, Rempe2019, Deng2024, Li2024, Pan2025}. Among all, dispersive coupling plays a promising role, as it can be effectively realized on traveling waves thanks to reflection of an optical pulse from a cavity containing a single $\Lambda$-type atom  in the limit of large cooperativity \cite{Wang2005,Rempe2019, Ulrik2022,Teja2023}, thus providing a method directly competing with Gaussian coupling protocols. 
Moreover, given the existing limitations in cavity QED experiments, it is crucial to design feasible deterministic protocols, that combine gates and operations already available with the state-of-the-art quantum optical technology and simultaneously offer scalable methods to extend the cavity QED capabilities, e.g. to target a wider class of QNG states for applications.

In this work, we propose a deterministic single-mode feedforward protocol based on dispersive coupling  [Fig.~\ref{fig:01-EvenGaussianSchemes}(a)], obtained by embedding the single-atom cavity in \cite{Rempe2019} with Gaussian electro-optical feedforward, with the advantages to reach %not only high average fidelity, but also required QNG interference fringes in phase space and displacement sensitivity.
both high average fidelity and required QNG interference fringes in phase space.
For a comparison, we briefly determine the performance achievable by Gaussian resources from nonlinear optics and highly efficient PNRDs by designing the heralded three-mode Gaussian-PNRD-based $\UG(2n)$ protocol, [Fig.~\ref{fig:01-EvenGaussianSchemes}(b)], and show the dispersive scheme to outperform it for large cat generation in essential quantum features.
%obtained by suitably optimized three-mode Gaussian coupling, 
%that provide a benchmark to demonstrate cavity QED advantage. Thereafter, we describe the dispersive-coupling protocol, obtained by embedding the single-atom cavity in \cite{Rempe2019} with Gaussian electro-optical feedforward, and address robustness of quantum non-Gaussian interference fringes against qubit decoherence and optical losses. Finally, we demonstrate application of the produced states for displacement sensing, i.e. weak force detection. 
The structure of the paper is the following. In Sec.s~\ref{sec:Proto} and~\ref{sec:Res} we describe the proposed $\UG(2n)$ and dispersive protocols and compare their performance in terms of fidelity and QNG features, respectively. Then, Sec.~\ref{sec:Imp} addresses the impact of imperfect dispersive coupling on the state preparation, while Sec.~\ref{sec:Disp} presents displacement sensitivity as a suitable metric for cat-state quality. Finally, in Sec.s~\ref{sec:Disc} and~\ref{sec:Concl}, we perform discussion about the obtained results and draw some conclusions, respectively.
Without loss of generality, we present the analysis by targeting the even cat $|\cat^{(+)}\rangle$, whereas the odd cat case is briefly discussed in~\cite{SuppMat}.

%%%%%%%%%%%%%%%%%%%%%%%%%%%%%%%%%%%%%%%%%%%%%%%%%%%%%%%%%
%%%%%%%%%%%%%%%%%%%%%%%%%%%%%%%%%%%%%%%%%%%%%%%%%%%%%%%%%
\section{The GP($2n$) and dispersive protocols}\label{sec:Proto}

%{\it The GP($2n$) and dispersive protocols.---} 
Upon generalization of photon addition and subtraction setups, we identify the scheme in Fig.~\ref{fig:01-EvenGaussianSchemes}(b) as the suitable configuration with the lowest number of modes for even-cat generation with Gaussian resources, hence referred to as the $\UG(2n)$ protocol. It is composed of an optimized Gaussian tritter, coupling vacuum modes $a,b,c$, %with two additional ones, $b$ and $c$
decomposed by Bloch-Messiah theorem into a sequence of pure single-mode squeezers $S(r_j)$ and passive operations \cite{Serafini2017,Houde2024}, after which
we introduce displacements $D_b(\beta)D_c(\beta')$, recently proved as necessary for quantum state engineering \cite{Impossibility}, and 
%the ancillary modes experience displacements $D_b(\beta)D_c(\beta')$, before being projected 
projection by PNRDs onto Fock states $|n\rangle$, thus heralding $2n$ photons in total. To guarantee even parity of the output state, it is crucial to choose a balanced second beam splitter with $\pi$ phase-shifts, i.e. $T'=1/2$, $\theta=\theta'=\pi$, and equal and opposite displacements amplitudes $\beta'=-\beta$, simplifying the tritter structure to the set of free parameters ${\bf p}=(r_1,r_2, r_3,T,\beta)$, where we further assume $r_j \in \mathbb{R}$ as $\alpha\ge 0$. We also note that for the choices 
$(r_2\!=\!r_3\!=\!0; \beta\!=\!0)$ and $(r_2\!=\!-r_1;r_3\!=\!0; T\!=\!1/2)$ we retrieve the subtraction and addition protocols respectively \cite{SuppMat}, justifying the $\UG(2n)$ as the more general lowest-mode Gaussian coupling scheme to achieve even parity.
%a unified framework for Gaussian coupling protocols \cite{Impossibility}. 
The $\bf p$-dependent quantum state is obtained as $|\phi_{2n}^{(+)} ({\bf p})\rangle= S(\chi) \sum_{k=0}^{n} d_{2k} |2k\rangle$ (derived in \cite{SuppMat}) that, by construction, gets finite stellar rank $\le 2n$, limited by maximal errorless $n$-photon resolution of the PRNDs. Then, we optimize ${\bf p}$ %the free parameters 
to maximize fidelity w.r.t. the target state, namely:
\begin{align}\label{eq:FideUG}
F^{(2n)}_{\UG}= \max_{ {\bf p} } \left| \left\langle \cat^{(+)} \Big|\phi_{2n}^{(+)} ({\bf p})\right\rangle\right|^2 \, .
\end{align}
We also discovered that, for the present case study, maximizing fidelity is equivalent to maximizing negativity of the $\UG(2n)$ Wigner function $W_\UG^{(2n)} (0,\bar{y})$ at the point $\bar{y}$ where the target Wigner function $W_{\rm cat}^{(+)}$ takes its absolute minimum \cite{Xanadu}.

To overcome the heralded $\UG(2n)$ scheme, we propose the %hybrid 
atom-light cavity QED protocol depicted in Fig.~\ref{fig:01-EvenGaussianSchemes}(a), where nonlinearity comes in the interaction and not only at detection stage. It extends the seminal idea firstly demonstrated by Hacker {\it et al.} in \cite{Rempe2019}, where a qubit and optical mode interact via dispersive Hamiltonian $H_{\DISP} = \theta \, \boldsymbol\sigma_{\!z} a^\dagger a$, with $\theta \ge0$, and $ \boldsymbol\sigma_{\!z}= |g\rangle \langle g| - |e\rangle \langle e|$ being the qubit Pauli $z$ operator \cite{Dispersive}, such that the unitary $U_\DISP=e^{-i H_\DISP}$ implements a conditional rotation of $a$ in either clockwise or anti-clockwise direction according to the qubit state. Initially, the dispersive protocol applies a $\pi/2$ pulse, namely $\theta= \pi/2$, onto coherent state of radiation $|i \alpha\rangle$, $\alpha\ge0$, and atomic state $|+\rangle= (|g\rangle+|e\rangle)/\sqrt{2}$, to create the ``atom-light entangled state" $|\Psi_{\DISP} \rangle =(|g\rangle |\alpha\rangle +  |e\rangle |-\alpha\rangle)/\sqrt{2}$ \cite{Rempe2019, Teja2023}, with subsequent qubit measurement in the $\{|\pm\rangle\}$ basis, such that, if $``+"$ is retrieved, light is projected exactly into the target state $|\cat ^{(+)}\rangle$, whereas in the opposite case we obtain the orthogonal odd cat. 

Now, to make the protocol deterministic, a suitable strategy to ``turn" the odd cat into the even one would be desirable. The problem has been addressed in circuit QED with different methods: (i) applying a subsquent conditional displacement gate $D(i\boldsymbol\sigma_{\!y} \beta)$ that approximately disentangle the qubit and light degrees of freedom in state $|\Psi_{\DISP} \rangle$ \cite{Liu2024, Singh2026}; (ii) inducing a non-adiabatic two-photon driving, available in Kerr nonlinear resonators \cite{Kerr1,Kerr2}; (iii) designing reservoir engineering schemes, that combine a classical \cite{Driving1} or two-photon \cite{Driving2, Driving3} driving with a suitable two-photon dissipation process.
We also remind that the Wigner function of a cat state exhibits a quantum interference pattern with Gaussian-shaped envelope between the two coherent-state peaks, with maximum (minimum) in the origin for the even (odd) state, see Fig.~\ref{fig:01-EvenGaussianSchemes}(a). Therefore, upon drawing inspiration on the displacement method in \cite{Driving1}, we can make our cavity QED protocol deterministic by using quantum optics feedforward techniques \cite{FF1, FF2,Notarnicola2023_HYNORE, Notarnicola_HY_PhN}. That is, when outcome ``-" is registered we apply the feedforward displacement operation $D( \pm i \gamma)$, $\gamma\ge0$ on the optical state, whose sign is randomly chosen with $50\%$ probability, inducing either positive or negative shift of the interference fringes along the $y$ direction of phase space. In this way we displace the odd cat to let its fringes overlap with those of the even state, and eventually obtain the mixture:
%Now, to make the protocol deterministic, when outcome ``-" is registered we apply the feedforward displacement operation $D( \pm i \gamma)$, $\gamma\ge0$ on the optical state, whose sign is randomly chosen with $50\%$ probability, inducing either positive or negative shift of the interference fringes along the $y$ direction of phase space, and eventually obtain the mixture:
\begin{align}\label{eq:rhoDISP}
\varrho_\DISP(\gamma) =&  p_+ |\cat ^{(+)}\rangle \langle \cat ^{(+)}| + \nonumber \\[1ex]
& p_- \left( \frac{|\psi'(\gamma)\rangle \langle\psi'(\gamma)| +|\psi'(-\gamma)\rangle \langle\psi'(-\gamma)|}{2} \right) \, ,
\end{align}
with $|\psi'(\gamma)\rangle=D(i \gamma)|\cat ^{(-)}\rangle$ and $p_\pm= (1\pm e^{-2\alpha^2})/2$ being the probability to detect the atom in the $|\pm\rangle$ state \cite{SuppMat}. The optimal displacement $\gamma$ is then chosen to maximize fidelity:
\begin{align}
F_\DISP&= \max_{\gamma} \, \left\langle \cat^{(+)} \right|\varrho_\DISP(\gamma) \left|  \cat^{(+)} \right\rangle \, .
\end{align}
Following the same approach, the dispersive protocol can be adopted to deterministically prepare the odd cat $|\cat^{(-)}\rangle$ with analogous performance to the even-cat case, see \cite{SuppMat}.

\begin{figure}
\includegraphics[width=0.95\columnwidth]{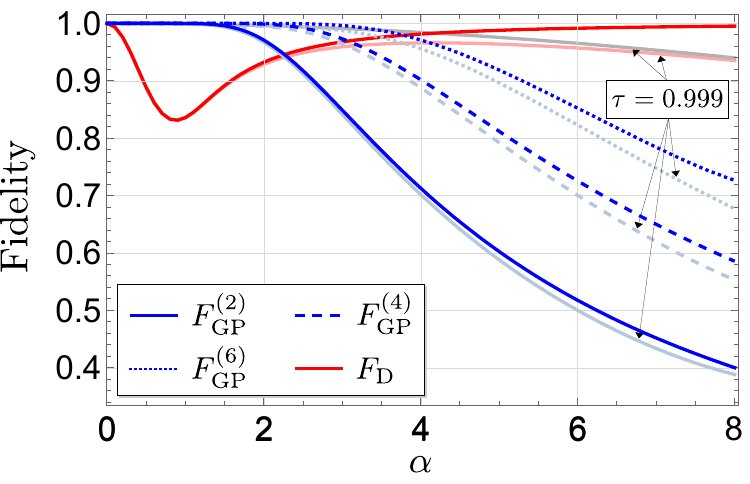}
\caption{Fidelities $F^{(2n)}_{\UG}$ for $n=1,2,3$ and $F_{\DISP}$ as a function of the target cat amplitude $\alpha$. Dispersive interaction provides a deterministic scheme that outperforms $\UG(2n)$ in the limit of large $\alpha$; in \cite{SuppMat} we discuss a probabilistic version of the protocol beating $\UG(2n)$ also for small $\alpha$. Moreover, the $\UG(2n)$ success probability for $\alpha\gg 1$ approaches $\approx 7.9\%, 3.0\%, 1.6\%$ for $n=1,2,3$ \cite{SuppMat}. The light-coloured lines represent fidelities after optical losses with transmissivity $\tau\le 1$, while the gray one refers to attenuation of the target cat $|\cat^{(+)}\rangle$.}\label{fig:02-Fidelity}
\end{figure}
\begin{figure*}
\includegraphics[width=0.95\linewidth]{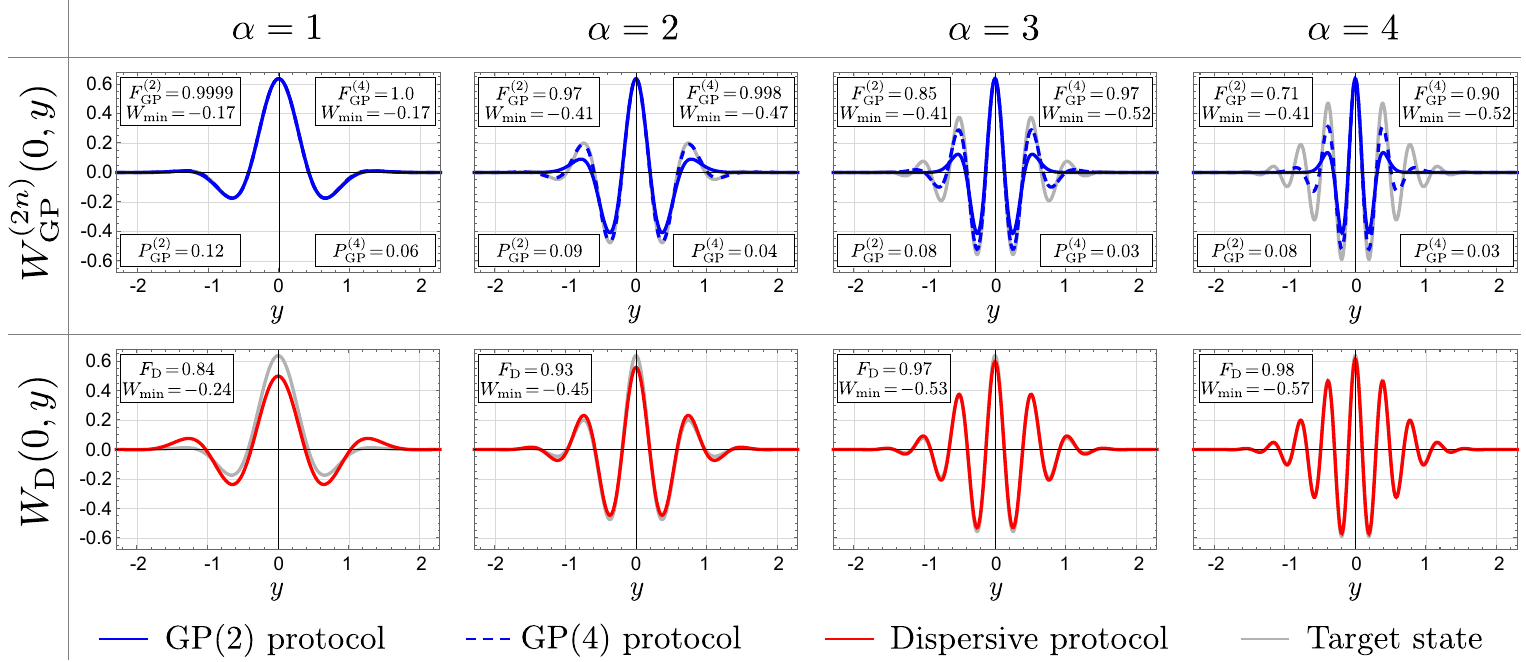}
\caption{Oscillations of the Wigner function along the $y$ axis for the $\UG(2n)$ protocol, with $n=1,2$ (top row) and the dispersive protocol (bottom row) for different values of the cat amplitude $\alpha$. Boxes report the corresponding values of fidelity and Wigner-function minima.}\label{fig:03-CutsWigner}
\end{figure*}

%%%%%%%%%%%%%%%%%%%%%%%%%%%%%%%%%%%%%%%%%%%%%%%%%%%%%%%%%
%%%%%%%%%%%%%%%%%%%%%%%%%%%%%%%%%%%%%%%%%%%%%%%%%%%%%%%%%
\section{Results}\label{sec:Res}

%{\it Results.---} 
Plots of fidelity and cuts of the Wigner function along the $y$ axis for both the $\UG(2n)$ protocol with $n=1,2,3$, and the dispersive scheme are reported in Fig.s~\ref{fig:02-Fidelity} and~\ref{fig:03-CutsWigner}, respectively. Due to the finite stellar rank, the $\UG(2n)$ performs well only in the kitten regime, while, for large enough $\alpha$, $F^{(2n)}_\UG$ becomes a decreasing function, and the interference pattern of the Wigner function exhibits only $2n$ negative dips, thus degrading the quality of the produced states for large $\alpha$. In this regime the success probability reaches $\approx 7.9\%, 3.0\%, 1.6\%$ for $n=1,2,3$ \cite{SuppMat}. On the contrary, the dispersive protocol achieves much better performance, as for $\alpha\gtrsim 0.8$ $F_\DISP$ becomes an increasing function that asymptotically approaches $1$ as:
\begin{align}\label{eq:AsymptoticDisp}
F_\DISP \approx 1 - \frac{\pi^2}{32 \alpha^2} \qquad \mbox{for $\alpha\gg 1$} \,,
\end{align}
where fidelity raises as a consequence of the feedforward displacement. Indeed, displacement of the odd-cat fringes makes states $D(\pm i\gamma)|\cat^{(-)}\rangle$ and $|\cat^{(+)}\rangle$ overlap, but worsens the correspondence between both their Gaussian envelopes and coherent-state peaks, so that, as the fringe spacing scales as $1/\alpha$ for $\alpha\gg 1$, the infidelity decreases as $1/\alpha^2$. Same asymptotic trend holds for the odd-cat generation \cite{SuppMat}.
Eq.~(\ref{eq:AsymptoticDisp}) provides comparable result to the protocol for microwave cats generation proposed in \cite{Liu2024, Singh2026} (whose fidelity scales as $\approx 1-\pi^2/64\alpha^2$), that, however, requires sequential iterations of dispersive couplings, whereas our proposal only needs a single light-atom interaction step, being assisted by Gaussian feedforward.
Besides, we observe many oscillations of the corresponding Wigner function, jointly appearing after only one interaction step, with much better accordance with the target state; minimum value of Wigner function equal to:
\begin{align}\label{eq:Wmin}
W_\DISP(0,\bar{y}) \approx W_{\rm cat}^{(+)} (0,\bar{y}) + \frac{\pi}{8\alpha^2}\qquad \mbox{for $\alpha\gg 1$} \,;
\end{align}
and associated distillable squeezing variance $V_\DISP \approx V_{\rm cat}^{(+)} - 1/8\alpha^4$ \cite{SuppMat,Distillation}, demonstrating dispersive interaction as the most powerful for large cat generation. In fact, while the $\UG(2n)$ conditionally projects a Gaussian state onto a finite superposition of Fock states of rank $\le 2n$, the dispersive protocol %is based on the superposition of unitary rotations. It 
exploits light-atom interaction to translate superposition created in the qubit to the optical mode, and experimental new part-Gaussian feedforward  to achieve determinism, thus combining
cavity QED experiments \cite{Wang2005,Rempe2019, Deng2024,Li2024, Pan2025, Ulrik2022,Teja2023} with well-established electro-optical feedforward methods \cite{FF1, FF2,Notarnicola2023_HYNORE, Notarnicola_HY_PhN}.

Fig.~\ref{fig:02-Fidelity} also shows robustness of the produced output states undergoing optical losses, modeled as a beam splitter of transmissivity $\tau \le 1$, in which case the dispersive fidelity $F_\DISP(\tau)$ loses the asymptotic behaviour~(\ref{eq:AsymptoticDisp}), becoming a decreasing function that for $\alpha\gg 1$ scales as:
\begin{align}
F_\DISP(\tau) \approx e^{-\alpha^2 \left(1-\sqrt{\tau}\right)^2} \frac{1+e^{-2\alpha^2(1-\tau)}}{2} \left( 1 - \frac{\pi^2}{32 \alpha^2} \right) \,,
\end{align}
whilst Wigner negativities are exponentially reduced \cite{SuppMat}.

%%%%%%%%%%%%%%%%%%%%%%%%%%%%%%%%%%%%%%%%%%%%%%%%%%%%%%%%%
%%%%%%%%%%%%%%%%%%%%%%%%%%%%%%%%%%%%%%%%%%%%%%%%%%%%%%%%%
\section{Robustness to imperfect dispersive coupling}\label{sec:Imp}

\begin{figure*}[t]
\includegraphics[width=0.99\linewidth]{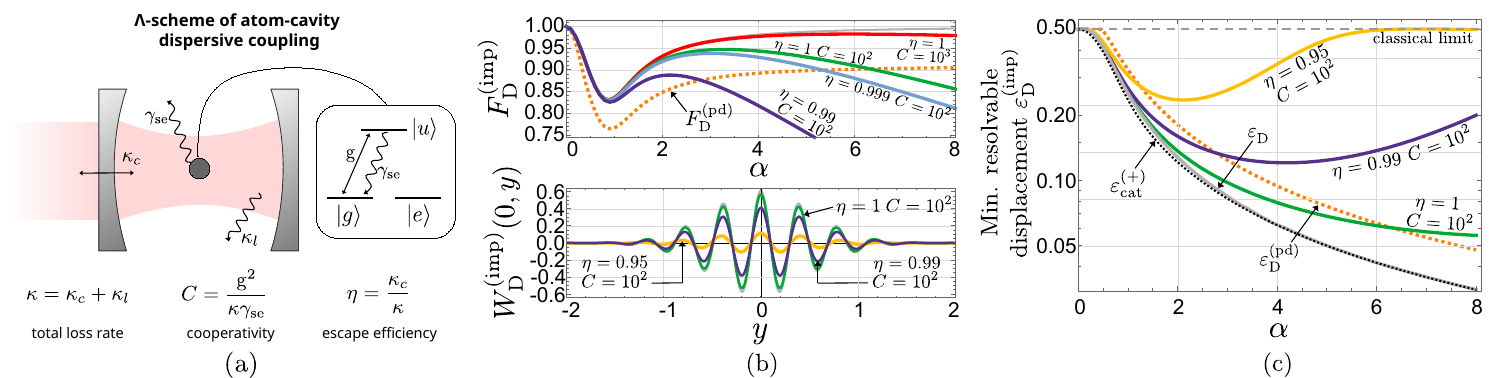}
\caption{(a) Cavity QED implementation of dispersive interaction, consisting of a cavity containing a $\Lambda$-type atom, coupled to impinging free-space field with rate $\kappa_c$ and to scattering and loss modes with rate $\kappa_l$. The two atomic levels $|g\rangle$ and $|u\rangle$ are resonantly coupled to the cavity field with gain $\rm g$, and transition $|g\rangle \to |u\rangle$ is also associated with spontaneously decay rate $\gamma_{\rm se}$ \cite{Rempe2019,Ulrik2022,Teja2023,SuppMat}. The system imperfections are conveniently described in terms of cooperativity $C$ and escape efficiency $\eta\le1$: ideal dispersive coupling is realized in the limits $C\gg1$ and $\eta=1$ (lossless cavity). (b) Fidelities $F_{\DISP}^{(\IMP)}$ (top) of the imperfect dispersive coupling protocol and Wigner interference fringes $W_\DISP^{(\IMP)}(0,y)$ with $\alpha=4$ (bottom) for different values of $C$ and $\eta$. The gray line corresponds to the ideal dispersive scheme; the dashed line refers to fidelity $F_{\DISP}^{(\PD)}$ in the presence of qubit phase damping of rate $\lambda=0.2$ \cite{SuppMat}. (c) Minimum resolvable displacements for ideal and imperfect dispersive schemes $\varepsilon_\DISP$ and $\varepsilon_\DISP^{(\IMP)}$ as a function of $\alpha$: the ideal protocol approaches the target state sensitivity  $\varepsilon_{\rm cat}^{(+)}$ for $\alpha\gg1$. The dashed line refers to $\varepsilon_{\DISP}^{(\PD)}$ in the case of qubit phase damping of rate $\lambda=0.2$ \cite{SuppMat}.
%cat states $|\cat^{(\pm)}\rangle$, equal to $\varepsilon_{\rm cat}^{(\pm)} = 1/2\sqrt{1+ 4\alpha^2/(1\pm e^{-2\alpha^2})}$. 
%The ideal protocol gets intermediate performance between $\varepsilon_{\rm cat}^{(-)}$ and $\varepsilon_{\rm cat}^{(+)}$, as state~(\ref{eq:rhoDISP}) is a convex combination of both even and displaced-odd cats. 
}\label{fig:04-Decoherence}
\end{figure*}

%{\it Robustness to imperfect dispersive coupling.---} 
The dispersive interaction required for the proposed protocol is implemented by the cavity QED setup in Fig.~\ref{fig:04-Decoherence}(a). To understand the difference to the $\UG(2n)$ scheme (mainly limited by PNRDs), we investigate its robustness to imperfect coupling realization, considering realistic conditions of finite cooperativity $C < \infty$ and non-unit escape efficiency $1/2<\eta<1$ \cite{Rempe2019, Ulrik2022, LachmannQUANTUM}. Both effects can be modeled as conditional optical losses with qubit-state dependent transmissivity $\eta_{g(e)}\le 1$, such that the ideally conditionally prepared states $|\cat^{(\pm)}\rangle$ decohere into the mixtures \cite{SuppMat}:
\begin{align}\label{eq:sigmaIMP}
\sigma_\pm^{(\IMP)} &= \frac{1}{4 p_{\pm}^{(\IMP)} } \Big\{ |\sqrt{\eta_g} \alpha\rangle\langle \sqrt{\eta_g} \alpha| +  |-\sqrt{\eta_e} \alpha\rangle\langle -\sqrt{\eta_e} \alpha| \nonumber \\ 
& \hspace{1.cm} \pm e^{-\Gamma \alpha^2/2} \Big(  |\sqrt{\eta_g} \alpha\rangle\langle- \sqrt{\eta_e} \alpha|+{\rm h.c.} \Big) \Big\} \, ,
\end{align}
with $\eta_g=[1-2\eta/(1+4C)]^2$, $\eta_e=(1-2\eta)^2$, $\Gamma=2-\eta_g-\eta_e+2\sqrt{(1-\eta_e)(\eta'-\eta_g)}$, $\eta'=1-16\eta C/(1+4C)^2$, and detection probabilities $p_{\pm}^{(\IMP)}=(1\pm e^{-[\Gamma+ (\sqrt{\eta_g}+\sqrt{\eta_e})^2]\alpha^2/2})/2$. 
The average output state becomes
$\varrho_\DISP^{(\IMP)}(\gamma) = p_+^{(\IMP)}  \sigma_+^{(\IMP)} +p_-^{(\IMP)} \sum_{{\rm k}=\pm} D( {\rm k}\,  i \gamma)  \sigma_-^{(\IMP)}  D\dag( {\rm k} \, i \gamma)/2 $,
%\begin{align}\label{eq:rhoIMP}
%\varrho_\DISP^{(\IMP)}(\gamma) = & p_+^{(\IMP)}  \sigma_+^{(\IMP)} + \nonumber \\& \hspace{.5cm} \frac{p_-^{(\IMP)}}{2} \sum_{{\rm k}=\pm} D( {\rm k}\,  i \gamma) %\, \sigma_-^{(\IMP)} \, D\dag( {\rm k} \, i \gamma)  \, ,
%\end{align}
with fidelity  $F_\DISP^{(\IMP)}= \max_{\gamma} \, \langle \cat^{(+)} |\varrho_\DISP^{(\IMP)}(\gamma)|  \cat^{(+)} \rangle$. As depicted in Fig.~\ref{fig:04-Decoherence}(b), for large $\alpha$, $F_\DISP^{(\IMP)}$ deviates gradually from the ideal case and becomes a decreasing function, thus expectedly losing the asymptotic trend, such that for $\alpha, C\gg 1$ and $1-\eta \ll 1$:
\begin{align}\label{eq:FIMP}
F_\DISP^{(\IMP)} \approx& \frac{1+e^{-2\alpha^2 \left(\frac{2-\eta}{4C}+1-\eta\right)}}{2} \, \left(1 - \frac{\pi^2}{32 \alpha^2}\right) \nonumber \\
& - \frac{\pi^2 (1-\eta)}{64 C} e^{-2\alpha^2 \left(\frac{2-\eta}{4C}+1-\eta\right)}  \, \left(1 - \frac{\pi^2}{16 \alpha^2}\right)  \, .
\end{align}
Instead, Wigner negativities, that witness QNG features, are not erased for any $\alpha$, albeit we observe $\alpha$-dependent suppression of oscillations, such that for $\alpha, C\gg 1$ and $1-\eta \ll 1$: 
\begin{align}\label{eq:WminIMP}
W_\DISP^{(\IMP)}(0,\bar{y})  \approx W_{\rm cat}^{(+)} (0,\bar{y}) +  \frac{f_1(\alpha)}{2\pi}  + \frac{\pi f_2(\alpha) }{8\alpha^2} \,,
\end{align}
with $f_1(\alpha)= 4-e^{-2\alpha^2(1-\eta+\eta/4C)}-3e^{-\alpha^2/2C}(1-\pi^2(1-\eta)/2C)$ and $f_2(\alpha)=-2+3e^{-\alpha^2/2C}(1-3\pi^2(1-\eta)/4C)$, whilst the distillable squeezing variance is increasing with $\alpha$, in faster way than negativity reduction \cite{SuppMat}.

With similar approach, in \cite{SuppMat} we further address the impact of decoherence before and inside qubit detection, by considering amplitude and phase damping noise, that, as reported in Fig.~\ref{fig:04-Decoherence}(b), induce only fidelity rescaling in the asymptotic limit and reduction of Wigner interference visibility proportional to the sole noise strength and independent of $\alpha$.

%%%%%%%%%%%%%%%%%%%%%%%%%%%%%%%%%%%%%%%%%%%%%%%%%%%%%%%%%
%%%%%%%%%%%%%%%%%%%%%%%%%%%%%%%%%%%%%%%%%%%%%%%%%%%%%%%%%
\section{Displacement sensitivity as interference measure}\label{sec:Disp}

%{\it Displacement sensitivity as interference measure.---} 
To further understand quality of the interference fringes, we test sensitivity of the previous quantum states to momentum displacement \cite{DispSensing_2002,DispSensing_2004}. The problem can be recast in the framework of quantum estimation theory \cite{Helstrom1976, Paris2009, Albarelli2020, NotarnicolaPhN, Notarnicola2024_Kerr}, where a weak displacement $D(i\varepsilon)=\exp(\sqrt{2} i \varepsilon \, \hat{x})$, $\varepsilon\ge0$ and $\hat{x}= (a+a\dag)/\sqrt{2}$ being the position-like operator, is applied to a probe state. Considering homodyne detection of quadrature $\hat{y}=i (a\dag-a)/\sqrt{2}$, the Cramér-Rao bound determines the minimum resolvable displacement $\varepsilon_{\rm min}= 1/\sqrt{\mathbb{F}}$ per experimental run, where
\begin{align}
\mathbb{F}=\int_\mathbb{R} dy \, \frac{1}{p(y+\sqrt{2}\varepsilon)} \left(\frac{d \,p(y+\sqrt{2}\varepsilon)}{d\varepsilon}  \right)^2
\end{align}
is the Fisher information (FI) associated with the momentum-like probability distribution $p(y)$ \cite{SuppMat}, %being also independent of $\epsilon$ as the problem is covariant , 
that for cat-state probes beats the classical limit, obtained with coherent states and corresponding to vacuum fluctuations, $\varepsilon_{\rm min}=1/2$. Fig.~\ref{fig:04-Decoherence}(c) shows
%compares the target cats sensitivity with 
the $\varepsilon_{\rm min}$ numerically computed for both the ideal and imperfect dispersive coupling protocols,
%from both states~(\ref{eq:rhoDISP}) and~(\ref{eq:rhoIMP}), 
proving the ideal coupling case to approach the target cat for $\alpha\gtrsim3$. Instead, imperfect coupling becomes detrimental in the large amplitude regime, where $\varepsilon_\DISP^{(\IMP)}$ increases with $\alpha$ and eventually equals the classical limit, whereas for $\alpha\ll 1$ we have $\varepsilon_\DISP^{(\IMP)} \approx \varepsilon_\DISP$. Therefore, error mitigation schemes \cite{LeJeannic2018,Teh2020} turn out to be essential also for high-quality deterministic state preparation. On the contrary, in the presence of qubit detection decoherence, $\varepsilon_{\rm min}$ remains a decreasing function for $\alpha\gg1$, thus maintaining the quantum advantage. Analogous results are also obtained for the quantum FI, that provides the optimal quantum-limited sensitivity, see \cite{SuppMat}. 
%More detailed discussion is reported in \cite{SuppMat}.

%%%%%%%%%%%%%%%%%%%%%%%%%%%%%%%%%%%%%%%%%%%%%%%%%%%%%%%%%
%%%%%%%%%%%%%%%%%%%%%%%%%%%%%%%%%%%%%%%%%%%%%%%%%%%%%%%%%
\section{Discussion}\label{sec:Disc}

The above results qualify feedforward operations as a powerful strategy to improve the performance of cavity QED experiments, for twofold reason. First, while conditional operations based on optical homodyne detection are common %procedure in various quantum information protocols 
\cite{Marek2018, Notarnicola2023_HFFRE,Suzuki2026}, hybrid atom-light feedforward in cavity QED has been only preliminary demonstrated in \cite{Hacker2016photon} to implement a photon-photon gate in a low-photon regime, and its potential has not been fully unveiled. Our results prove atom-light feedforward as crucial to achieve determinism in large cat state preparation, and provide a cornerstone towards further extension of the cavity QED capabilities. For instance, by sequential application of the proposed dispersive gate to two subsequent coherent pulses $|\alpha\rangle |\beta\rangle$, followed by atom detection in the $|\pm\rangle$ basis and double feedforward displacement, one can target the nonlocal cat states $|\alpha\rangle |\beta\rangle \pm |-\alpha\rangle |-\beta\rangle$ and achieve deterministic entanglement generation, being resource for quantum error correction \cite{Hastrup2022}.

Secondly, adopting electro-optical feedforward may simplify deterministic atom-light coupling protocols. In fact, the existing schemes for cat preparation require two subsequent conditional displacement gates $D(\boldsymbol\sigma_{\!x} \alpha) D(i\boldsymbol\sigma_{\!y} \beta)$ \cite{HybridQuantum,Kikura2026}, that %, with $\beta=\pi/8\alpha$, that 
%can be implemented by either combination of multiple displacements and dispersive gates \cite{HybridQuantum}, or by a suitably engineered optical cavity containing a detuned four-level atom \cite{Kikura2026}. In both cases, this 
require multiple reflections of the optical beam to the cavity, and precise control of the experimental setup in the realistic conditions of finite cooperativity and non-unit escape efficiency. In contrast, our protocol replaces gate $D(i\boldsymbol\sigma_{\!y} \beta)$ with Gaussian feedforward to reduce the setup complexity to one single cavity reflection, obtaining comparable performance to circuit QED schemes \cite{Liu2024, Singh2026}. 
Together with setup simplification, this choice also paves the way for distributed quantum experiments, where quantum gates can be obtained via nonlocal interaction of two remote subsystems \cite{NonlocalGates}.
While the current demonstrations of nonlocal gates have been only carried out in discrete-variable platforms \cite{Nonlocal1, Nonlocal2, Nonlocal3}, our results offer a powerful strategy to extend this methodology to hybrid platforms, with the near-future objective of remote hybrid quantum networks.

%. Such hybrid gates can be implemented by either combination of multiple displacements and dispersive gates \cite{} UD= exp(-i θ a†a) \cite{HybridQuantum}, or by an optical cavity that contains a suitably engineered four-level atom recently proposed in \cite{Kikura2026}. In both cases, those protocols require two hybrid gates, and, consequently, multiple reflections of optical beam to the cavity. While this would be feasible for microwaves (circuit QED experiments), it represents a challenging task for cavity QED at optical frequencies, which needs precise control of the experimental setup in the realistic conditions of finite cooperativity and non-unit escape efficiency to avoid accumulation of unwanted errors. 
%For both reasons, feedforward-assisted cavity QED experiments candidate themselves as a natural platform for cat state generation, with high impact for future realizations of nonGaussian quantum information protocols. 

%%%%%%%%%%%%%%%%%%%%%%%%%%%%%%%%%%%%%%%%%%%%%%%%%%%%%%%%%
%%%%%%%%%%%%%%%%%%%%%%%%%%%%%%%%%%%%%%%%%%%%%%%%%%%%%%%%%
\section{Conclusions}\label{sec:Concl}

%{\it Conclusions.---} 
In this work, we proposed a deterministic feedforward-based protocol using dispersive interaction with a qubit for cat state generation, proving it to outperform the Gaussian-PNRD-based protocol for $\alpha\gg 1$. We demonstrate robustness of its QNG features against impefect coupling and qubit decoherence, showing dispersive coupling as promising for the following reasons: determinism, large fidelity, high-visibility quantum interference obtained after a single interaction step, and flexible implementation by cavity-reflection of coherent traveling beams. Our results also prove FI, already proposed as a quantum macroscopicity measure \cite{Macroscopic2015,Macroscopic2016,Macroscopic2017,Macroscopic2018,Macroscopic2019,Macroscopic2025}, as a reliable witness of cat-state interference. In fact, while fidelity and Wigner negativity are exponentially decreasing for $\alpha\gg1$ in the presence of noise, the minimum resolvable displacement provides a more robust quantifier of large cat-state quality, and reveals its nonclassicality also in the suboptimal, but feasible, case of homodyne detection.
Moreover, our scheme can be straightforwardly extended to target a wider class of superposition states, e.g. (i) unbalanced cat states $|\psi\rangle \propto c_1 |\alpha\rangle + c_{2} |-\alpha\rangle$ relevant for phase sensing, by preparing the initial qubit state into the superposition $c_{1} |g\rangle + c_{2} |e\rangle$; (ii) squeezed cat states $S(r) |\cat^{(\pm)}\rangle$, $r\in \mathbb{R}$, obtained with the same fidelity depicted in Fig.~\ref{fig:02-Fidelity}
by considering an input squeezed coherent state $S(-r) |i \alpha\rangle$ and the feedforward Gaussian operation $S(r) D(\gamma) S(-r) = D(e^r \gamma)$; (iii) multi-headed cat states by subsequent beam reflection from array cavities; (iv) multimode and nonlocal cats.
The proposed protocol can be also translated to other platforms, including microwaves \cite{Girvin2019, He2023,Milul2023, Hot2025,Hutin2025}, atomic ensembles \cite{Nori2021,Omran2019} and mechanical modes \cite{QuantumDot, Kanari2022,Fadel, MechQubit}, thus providing a crucial step to target exotic states of radiation for quantum information purposes.
\textcolor{white}{\nocite{BCH, Olivares2021, Jaromir1, Jaromir2, Dolinar1973, Sych2016, Izumi2013, Notarnicola2023_KOR, NotarnicolaTesi, Hage2007, Marek2007, DarrenDSQ, Fiurasek2025, Breuer2002, Collett1984, Wiseman2009, NielsenChuang}}

%
%, lead to the generation of multi-headed cat states, e.g. the four-component state $|\psi\rangle \propto |\alpha\rangle + |i \alpha\rangle + |-\alpha\rangle + |-i\alpha\rangle$.
%
%On the other hand, we proved dispersive protocol to be more fragile against to optical losses, especially in the asymptotic limit $\alpha \gg 1$. To this aim, it is crucial to protect the optical mode from delay-line decoherence, e.g. by exploiting optical parametric amplification, that has already been proved useful for loss mitigation is different contexts \cite{LeJeannic2018,Kalash2023, Lukanowski2023, Notarnicola_MultiSpan}.

%%%%%%%%%%%%%%%%%%%%%%%%%%%%%%%%%%%%%%%%%%%%%%%%%%%%%%%%%
%%%%%%%%%%%%%%%%%%%%%%%%%%%%%%%%%%%%%%%%%%%%%%%%%%%%%%%%%
\section*{Acknowledgements}

%{\it Acknowledgements.---} 
The authors thank J.~Fiurá\v{s}ek for stimulating discussion about $\UG(2n)$ scheme. M.N.N. acknowledges the QuantERA project CLUSSTAR (8C2024003) of the MEYS of the Czech Republic. Project CLUSSTAR has received funding from the EU Horizon Programme under Grant Agreement No. 731473 and 101017733 (QuantERA). R.F. acknowledges the project 23-06308S of the Czech Science Foundation, the EU Horizon Programme under Grant Agreement No.101080173 (CLUSTEC), No. 101257526 (SUPERSPIN) and the project CZ$.02.01.010022\_0080004649$ (QUEENTEC) of the EU and the Czech Ministry of Education, Youth and Sport.

%%%%%%%%%%%%%%%%%%%%%%%%%%%%%%%%%%%%%%%%%%%%%%%%%%%%%%%%%
%%%%%%%%%%%%%%%%%%%%%%%%%%%%%%%%%%%%%%%%%%%%%%%%%%%%%%%%%

\bibliography{BiblioCatStates.bib}

%%%%%%%%%%%%%%%%%%%%%%%%%%%%%%%%%%%%%%%%%%%%%%%%%%%%%%%%%
%%%%%%%%%%%%%%%%%%%%%%%%%%%%%%%%%%%%%%%%%%%%%%%%%%%%%%%%%

\end{document}

\title{Deterministic feedforward-based generation of large optical coherent-state superposition. \\Supplemental material}
\author{Michele N.~Notarnicola}
%\email{michelenicola.notarnicola@upol.cz}
\affiliation{Department of Optics, Palack\'y University,
17. Listopadu 12, 779 00 Olomouc (Czech Republic) }

\author{Marcin Jarzyna}
\affiliation{Centre for Quantum Optical Technologies, 
Centre of New Technologies, 
University of Warsaw, Banacha 2c, 02-097 Warszawa (Poland)}

\author{Radim Filip}
\affiliation{Department of Optics, Palack\'y University,
17. Listopadu 12, 779 00 Olomouc (Czech Republic) }
\date{\today}
%%%%%%%%%%%%%%
\begin{abstract}

\end{abstract}
\maketitle
%%%

\begin{figure}
\includegraphics[width=0.99\columnwidth]{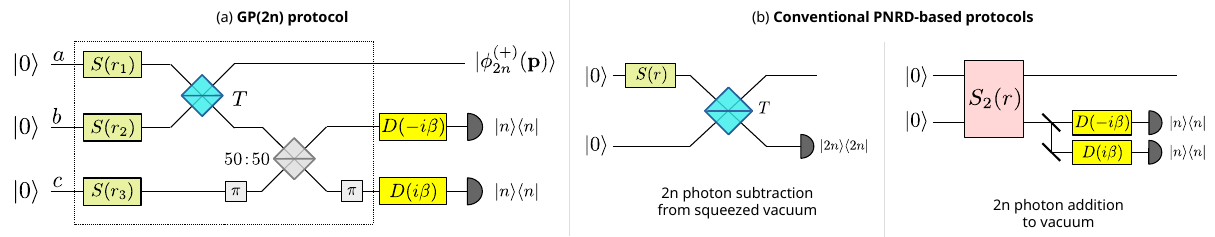}
\caption{(a) Gaussian-PNRD-based [$\UG(2n)$] protocol for even cat generation, performing conditional displacement-PNRD over an optimized three-mode Gaussian state. The choices of both a balanced second beam splitter with $\pi$ phase-shifts and equal and opposite imaginary displacements amplitudes, are essential to obtain even parity of the resulting output state $|\phi_{2n}^{(+)} ({\bf p})\rangle$. (b) Conventional methods for cat generation employing PNRDs, namely $2n$ photon-subtraction from squeezed vacuum and $2n$ photon-addition to vacuum.}\label{fig:S01-GP2nSchemes}
\end{figure}
%--------------------------------------------------------------------------------------------------------------------------------------------------%
\section{Detailed derivation of the $\UG(2n)$ protocol}\label{sec:GP2n}
In this Section, we present the extended calculations to derive the conditional quantum state $|\phi_{2n}^{(+)} ({\bf p})\rangle$ associated with the $\UG(2n)$ protocol. We start by observing that the $\UG$ setup schematized in Fig.~\ref{fig:S01-GP2nSchemes}(a) is associated with the Kraus operator:
\begin{align}
X_{2n} &= {}_{b}\langle n|{}_{c}\langle n| \, \Big\{ D_b(-i\beta) D_c(i\beta) U_{cb}^{(\rm BS)}(1/2)U_{ab}^{(\rm BS)}(T) \Big\} |r_2\rangle_b |r_3\rangle_{c} \,\, S_a(r_1)\nonumber \\[1ex]
&= \frac{1}{n!}{}_{bc}\langle 00| \, b^n c^n \, \Big\{D_b(-i\beta) D_c(i\beta) U_{cb}^{(\rm BS)}(1/2) U_{ab}^{(\rm BS)}(T) \Big\} |r_2, r_3\rangle_{bc}  \,\,  S_a(r_1)  \, ,
\end{align}
where $S_{a_j}(r_j)=\exp\{r_j [({a_j}\dag)^2-{a_j}^2]/2\}$, $r_j \in \mathbb{R}$, $a_j=a,b,c$, is the squeezing operator, $|r_2\rangle_b= S_b(r_2)|0\rangle_b$, $|r_3\rangle_c= S_c(r_3)|0\rangle_b$, $U_{ab}^{(\rm BS)}(T)=\exp[ \arccos(\sqrt{T}) (a\dag b - a b\dag)]$ describes beam splitter interaction, and $D_b(-i \beta)= \exp[ - i \beta (b\dag +b)]$, $D_c(i \beta)= \exp[ i \beta (c\dag +c)]$ are displacements, where the amplitudes have been conveniently re-defined as $\mp i \beta$, with $\beta^2 \in \mathbb{R}$, respectively. We also note that the $\pi$ phase-shifts in the last mode of the tritter act as a swap between $b$ and $c$ in the second beam splitter operator, that reads $U_{cb}^{(\rm BS)}(1/2)=\exp[ \pi (c\dag b - c b\dag)/4]$.

By exploiting the Heisenberg evolution of modes $b$ and $c$, we rewrite the previous expression as:
\begin{align}\label{eq:X2nA}
X_{2n} &= \frac{1}{n!} \, {\textcolor{white}{\Big|}}_{bc}\left\langle 0, -\sqrt{2} i \beta\right| \left(\frac{b-c}{\sqrt{2}}-i\beta\right)^n\left(\frac{c+b}{\sqrt{2}}+i\beta\right)^n \, U_{ab}^{(\rm BS)}(T) \Big|r_2, r_3\Big\rangle_{bc}   \,\,  S_a(r_1) \nonumber \\[1ex]
&= \frac{1}{2^n n!} \, {\textcolor{white}{\Big|}}_{bc}\left\langle 0, -\sqrt{2} i \beta\right|  \left[b^2-\left(c+\sqrt{2} i\beta\right)^2 \right]^n \, U_{ab}^{(\rm BS)}(T)  \Big|r_2, r_3\Big\rangle_{bc}  \,\,  S_a(r_1) \nonumber  \\[1ex]
&= \frac{1}{2^n n!} \, \sum_{j=0}^n \binom{n}{j} (-1)^{j} \, \langle -\sqrt{2} i \beta | (c+\sqrt{2} i\beta)^{2j} |r_3\rangle \,\times \, {}_{b}\langle 0| \,  b^{2(n-j)} \, U_{ab}^{(\rm BS)}(T)  |r_2\rangle_{b}  \,\,  S_a(r_1) \, ,
\end{align}
$|-\sqrt{2}i\beta\rangle_c$ being a coherent state with amplitude $-\sqrt{2}i\beta$. We underline that the structure of Eq.~(\ref{eq:X2nA}) justifies the need of a balanced second beam splitter to achieve even parity of the resulting output state, as, after application of $U_{ab}^{(\rm BS)}(T)$ and partial trace over $b$, only even powers of the signal mode $a$ will be present. 
%
%&= \frac{1}{2^n n!} \, {}_{bc}\left\langle 0, -\sqrt{2} i \beta\right|  U_{ab}^{(\rm BS)}(T) \left[\left(\sqrt{T} b-\sqrt{1-T} a \right)^2-\left(c+\sqrt{2} i\beta\right)^2 \right]^n \Big|r_2, r_3\Big\rangle_{bc}  \,\,  S_a(r_1) \, ,

We now start considering the simplest case $n=1$, yielding the $\UG(2)$ protocol. First of all, we perform useful preliminary calculations. By resorting to the Fock-state expansion $|r_3\rangle= \mu_3^{-1/2} \sum_n (\lambda_3/2)^n \sqrt{(2n)!}/n! |n\rangle$, with $\mu_3=\cosh(r_3)$ and $\lambda_3=\tanh(r_3)$, we have:
\begin{align}\label{eq:ck}
\langle - \sqrt{2} i \beta | c^k | r_3\rangle &= \frac{1}{\sqrt{\mu_3}} \sum_{n=0}^\infty \left(\frac{\lambda_3}{2}\right)^n \frac{\sqrt{(2n)!}}{n!} \langle - \sqrt{2} i \beta | c^k | 2n\rangle \nonumber  \\[1ex]
&= \frac{e^{-|\beta|^2}}{\sqrt{\mu_3}}  \sum_{n=0}^\infty \left(\frac{\lambda_3}{2}\right)^n (\sqrt{2} i \beta^*)^{2n-k} \frac{(2n)!}{n! (2n-k)!} \, , \qquad k\in \mathbb{N} \, .
\end{align}
%
%\begin{subequations}
% \begin{align}
%Y_2 &= \langle -\sqrt{2} i \beta |0\rangle \,\, {}_{b} \langle 0 \, |\,  U_{ab}^{(\rm BS)}(T)  \,\, \times \nonumber \\[1ex]
%& \hspace{1.5cm} \Big\{ \frac12 \left(\sqrt{T} b -\sqrt{1-T} a\right)^2 + \beta^2 \Big\} |r_2\rangle_{b}  \\[1ex]
%&= \frac{e^{-\beta^2}}{2} \Big\{ \left[(1-T) a^2 + 2 \beta^2\right] \, {}_b\langle 0| U_{ab}^{(\rm BS)}(T) |r_2\rangle_{b}  \nonumber \\[1ex]
%& \hspace{1.5cm} - 2 a \sqrt{T(1-T)} \, {}_b\langle 0| \, U_{ab}^{(\rm BS)}(T)b  |r_2\rangle_{b}   \nonumber \\[1ex]
%& \hspace{3.cm}+ T \,\, {}_b\langle 0| U_{ab}^{(\rm BS)}(T) b^2|r_2\rangle_{b} \Big\} \, . \label{eq:Y_2final}
%\end{align}
%\end{subequations}
Then, we invoke the Baker-Campbell-Hausdorff (BCH) decomposition of the beam splitter unitary operator, namely $U_{ab}^{(\rm BS)}(T)= e^{-\sqrt{(1-T)/T} a b\dag} (\sqrt{T})^{a\dag a - b\dag b} e^{\sqrt{(1-T)/T} a\dag b}$ \cite{BCH, Olivares2021}, to evaluate the quantities:
\begin{subequations}\label{eq:langlerangle}
 \begin{align}
&{}_b\langle 0| U_{ab}^{(\rm BS)}(T) |r_2\rangle_{b} = \frac{1}{\sqrt{\mu_2}} (\sqrt{T})^{a\dag a} e^{\frac{1-T}{2T} \, \lambda_2 (a\dag)^2} \, ,
\end{align}
 \begin{align}
&{}_b\langle 0| \, U_{ab}^{(\rm BS)}(T)b  |r_2\rangle_{b} = \frac{ \lambda_2}{\sqrt{\mu_2}} (\sqrt{T})^{a\dag a} \, \left( \sqrt{\frac{1-T}{T}} a\dag \right) e^{\frac{1-T}{2T} \, \lambda_2 (a\dag)^2} \, ,
\end{align}
\begin{align}
&{}_b\langle 0| U_{ab}^{(\rm BS)}(T) b^2|r_2\rangle_b = \frac{\lambda_2 }{\sqrt{\mu_2}} (\sqrt{T})^{a\dag a}    \, \left[1+ \frac{1-T}{T} \lambda_2 (a\dag)^2 \right] e^{\frac{1-T}{2T} \, \lambda_2 (a\dag)^2} \, ,
\end{align}
\end{subequations}
with $\lambda_2=\tanh(r_2)$ and $\mu_2=\cosh(r_2)$. By plugging Eq.s~(\ref{eq:ck}) and~(\ref{eq:langlerangle}) into~(\ref{eq:X2nA}), and also resorting to the BCH formula for single mode squeezing $S_a(r_1)= e^{\lambda_1 (a\dag)^2/2} \, \mu_1^{-a\dag a -1/2} e^{-\lambda_1 a^2/2}$, where $\lambda_1=\tanh(r_1)$ and $\mu_1=\cosh(r_1)$ \cite{BCH, Olivares2021}, we eventually obtain $X_2$ in the fascinating expression:
\begin{align}\label{eq:X2finalexp}
X_2 &= \frac{e^{-|\beta|^2 -\lambda_3(\beta^*)^2}}{2\sqrt{\mu_1 \mu_2 \mu_3}}  \left\{ {\cal C}_2^{(2)} \left(a-\lambda_2 a\dag\right)^2 +{\cal C}_0^{(2)} \right\} \, \, e^{\frac{\xi}{2} (a\dag)^2} \left( \frac{\sqrt{T}}{\mu_1} \right)^{a\dag a} e^{-\frac{\lambda_1}{2} a^2} \, ,
\end{align}
with: 
\begin{align}
{\cal C}_2^{(2)}= \frac{1-T}{T} \qquad \mbox{and} \qquad {\cal C}_0^{(2)}=  \lambda_2  -\lambda_3 +2 \beta^2 (1+\lambda_3)^2\, ,
\end{align}
%$c_1=(1-T)/T$, $c_2=  \lambda_2  -\lambda_3 +2 \beta^2 (1+\lambda_3)^2$, 
and
\begin{align}
\xi= T \lambda_1 + (1-T) \lambda_2 \, .
\end{align}
Accordingly, we derive the not-normalized output state $|\widetilde{\phi}_{2}^{(+)} ({\bf p})\rangle = X_2 |0\rangle_a$ as:
\begin{align}
|\widetilde{\phi}_{2}^{(+)} ({\bf p})\rangle= \frac{e^{-|\beta|^2 -\lambda_3(\beta^*)^2}}{2\sqrt{\mu_1 \mu_2 \mu_3}}  (1-\xi^2)^{-1/4} \,\, |\widetilde{\psi}_2({\bf p})\rangle \, ,
\end{align}
where:
\begin{align}
|\widetilde{\psi}_2({\bf p})\rangle= \left\{ {\cal C}_2^{(2)} \left(a-\lambda_2 a\dag\right)^2 + {\cal C}_0^{(2)}\right\} |\chi\rangle \, ,
\end{align}
and $|\chi\rangle=S_a(\chi)|0\rangle$ is a single-mode squeezed state with squeezing parameter $\chi=\tanh^{-1}(\xi)$. The corresponding success probability reads:
\begin{align}
P_2({\bf p})= \frac{e^{-2|\beta|^2 -2\lambda_3{\rm Re}[(\beta^*)^2]}}{4\mu_1 \mu_2 \mu_3}  (1-\xi^2)^{-1/2} \, {\cal K}_2({\bf p}) \, ,
\end{align}
${\cal K}_2({\bf p})= \langle \widetilde{\psi}_2({\bf p})|\widetilde{\psi}_2({\bf p})\rangle$ being a normalization factor, while the output normalized state becomes:
\begin{align}\label{eq:phi2}
|\phi_2^{(+)} ({\bf p})\rangle&= \frac{ |\widetilde{\psi}_2({\bf p})\rangle}{\sqrt{{\cal K}_2({\bf p})}} =\frac{1}{\sqrt{{\cal K}_2({\bf p})}} \left\{ {\cal C}_2^{(2)} \left(a-\lambda_2 a\dag\right)^2 +{\cal C}_0^{(2)} \right\} |\chi\rangle \, ,
\end{align}
being a linear combination of the (Gaussian) state $|\chi\rangle$ and the photon-subtracted and/or added squeezed states, $a^2|\chi\rangle$, $(a\dag)^2|\chi\rangle$ and $a\dag a|\chi\rangle$, that consequently gets stellar rank $\le 2$. 

With analogous procedure and more cumbersome calculations, we retrieve the general expression of the Kraus operator for $n>1$ as:
\begin{align}\label{eq:X4finalexp}
X_{2n} &= \frac{e^{-|\beta|^2 -\lambda_3(\beta^*)^2}}{2^n n!\sqrt{\mu_1 \mu_2 \mu_3}}  \left\{ \sum_{k=0}^n {\cal C}_{2k}^{(2n)} \left(a-\lambda_2 a\dag\right)^{2n}  \right\} \, \, e^{\frac{\xi}{2} (a\dag)^2} \left( \frac{\sqrt{T}}{\mu_1} \right)^{a\dag a} e^{-\frac{\lambda_1}{2} a^2} \, ,
\end{align}
with corresponding success probability and output state:
\begin{align}\label{eq:SRATE}
P_{2n}({\bf p})= \frac{e^{-2|\beta|^2 -2\lambda_3{\rm Re}[(\beta^*)^2]}}{(2^n n!)^2 \mu_1 \mu_2 \mu_3}  (1-\xi^2)^{-1/2} \, {\cal K}_{2n}({\bf p}) \qquad \mbox{and} \qquad
|\phi_{2n}^{(+)} ({\bf p})\rangle= \frac{ |\widetilde{\psi}_{2n}({\bf p})\rangle}{\sqrt{{\cal K}_{2n}({\bf p})}} \, ,
\end{align}
where $|\widetilde{\psi}_{2n}({\bf p})\rangle=\sum_{k=0}^n {\cal C}_{2k}^{(2n)} (a-\lambda_2 a\dag)^{2n} |\chi\rangle $ and ${\cal K}_{2n}({\bf p})= \langle \widetilde{\psi}_{2n}({\bf p})|\widetilde{\psi}_{2n}({\bf p})\rangle$.

However, we did not find a recursive relation to explicitly compute the coefficients ${\cal C}_{2k}^{(2n)}$, therefore one should address each case study separately. In particular, for $n=2$ we have:
 \begin{align}
{\cal C}_4^{(4)}&= \left(\frac{1-T}{T}\right)^2 \, , \nonumber \\ 
{\cal C}_2^{(4)}&= \frac{1-T}{T}  \left(6\lambda_2 -2\lambda_3+4\tilde{\beta}^2\right) \, ,\nonumber \\[1ex]
{\cal C}_0^{(4)}&= 3 \lambda_2^2 -2 \lambda_2\left(\lambda_3-2\tilde{\beta}^2\right)+3\lambda_3^2-12 \lambda_3\tilde{\beta}^2 + 4\tilde{\beta}^4 \, ,
\end{align}
and for $n=3$:
 \begin{align}
{\cal C}_6^{(6)}&= \left(\frac{1-T}{T}\right)^3 \, ,\nonumber \\
{\cal C}_4^{(6)}&= 3\left(\frac{1-T}{T}\right)^2  \left(5\lambda_2 -\lambda_3+2\tilde{\beta}^2\right) \, ,\nonumber \\[1ex]
{\cal C}_2^{(6)}&= 3\left(\frac{1-T}{T}\right) \Big\{ 15 \lambda_2^2 + 3 \lambda_3^2-12\lambda_3 \tilde{\beta}^2 + 4\tilde{\beta}^4-6\lambda_2 \left(\lambda_3-2 \tilde{\beta}^2\right)\Big\} \, ,\nonumber \\
{\cal C}_0^{(6)}&= 15 \lambda_2^3-15\lambda_3^3+90\lambda_3^2\tilde{\beta}^2-60\lambda_3 \tilde{\beta}^4+8 \tilde{\beta}^6 - 9\lambda_2^2 \left(\lambda_3-2 \tilde{\beta}^2\right)+3\lambda_2 \left( 3 \lambda_3^2-12\lambda_3 \tilde{\beta}^2 + 4\tilde{\beta}^4\right)\, ,
\end{align}
where $\tilde{\beta}=\beta(1+\lambda_3)$.
The $\UG(2n)$ protocol is then obtained by determining the parameters ${\bf p}$ that maximize fidelity with the target state, namely $
F_\UG^{(2n)}= \max_{{\bf p}} | \langle \cat^{(+)} |\phi_{2n}^{(+)} ({\bf p})\rangle|^2$.
To do so, we exploit the recent techniques developed in \cite{Jaromir1,Jaromir2}. As a convenient computational trick, we formally re-express the normalized state $|\phi_{2n}^{(+)} ({\bf p})\rangle$ as a squeezed core state of stellar rank $2n$, that is:
\begin{align}\label{eq:Core}
|\phi_{2n}^{(+)} ({\bf p})\rangle = S(\chi) \sum_{k=0}^{n} d_{2k} |2k\rangle \, ,
\end{align}
with coefficients $d_{2k}\ge0$ related to the actual physical parameters in Fig.~\ref{fig:S01-GP2nSchemes}(a) and the  $\{ {\cal C}_{2k}^{(2n)}\}$ in Eq.~(\ref{eq:X4finalexp}) by: 
\begin{align}\label{eq:dk}
d_{2k}= \frac{1}{\sqrt{{\cal K}_{2n}({\bf p})}} \sum_{j=k}^n \frac{ {\cal C}_{2j}^{(2n)}(2j)!}{2^{j-k} (j-k)! \sqrt{(2k)!}} \frac{(1-\lambda_2 \xi)^{j-k} (\lambda_2-\xi)^{j+k}}{(1-\xi^2)^j}  \,, \quad k=0,\ldots, n-1 \, ,
\end{align}
and $d_{2n}= \sqrt{1-\sum_{k=0}^{n-1}(d_{2k})^2}$. In light of this, the normalization factor is also found as:
\begin{align}
{\cal K}_{2n}({\bf p})=\sum_{k=0}^{n} \left(\sum_{j=k}^n \frac{ {\cal C}_{2j}^{(2n)}(2j)!}{2^{j-k} (j-k)! \sqrt{(2k)!}} \frac{(1-\lambda_2 \xi)^{j-k} (\lambda_2-\xi)^{j+k}}{(1-\xi^2)^j}\right)^2 \, .
\end{align}
Then, we proceed in three steps:
\begin{enumerate}[(i)]
\item First of all, we find the optimal set of $n+1$ coefficients $\{\chi^{(2n)}, d_0^{(2n)},\ldots,d_{2n-2}^{(2n)}\}$ that maximize fidelity of a quantum state with the structure~(\ref{eq:Core}) w.r.t. the target even cat. 
\item Then, we interpret Eq.~(\ref{eq:dk}) plus condition $\xi= \tanh(\chi)= T \lambda_1 + (1-T) \lambda_2$ as a (nonlinear) algebraic system of $n+1$ constraints for the 5 physical setup parameters ${\bf p}=(r_1,r_2,r_3,T,\beta)$ of the three-mode Gaussian state involved in the $\UG(2n)$ protocol. That is, we invert~(\ref{eq:dk}) to derive a subset of $n+1$ of the 5 physical parameters as a function of the optimal values $\{\chi^{(2n)}, d_0^{(2n)},\ldots,d_{2n-2}^{(2n)}\}$ numerically derived in the previous step. 
\item Finally, the remaining $5-n-1$ physical parameters provide free variables that do not affect the shape of the output state $|\phi_{2n}^{(+)} ({\bf p})\rangle$, and can be eventually optimized to only maximize the success probability $P_{2n} ({\bf p})$ of Eq.~(\ref{eq:SRATE}).
\end{enumerate}
In the end, we obtain a set of optimized values ${\bf p}_{2n}=(r_{1}^{(2n)}, r_{2}^{(2n)}, r_{3}^{(2n)}, T^{(2n)}, \beta^{(2n)}) $, that maximize fidelity $F_\UG^{(2n)}$, with the corresponding success rate
\begin{align}
P_{\UG}^{(2n)}= P_{2n} \left({\bf p}_{2n} \right) \, .
\end{align}
The previous argument also proves that the $\UG(2n)$ scheme is optimal for preparation of cat-like states with stellar rank $2n \le 8$, as the procedure of step (ii) involves solving a system of $n+1$ equations with $5$ variables (corresponding to the physical parameters involved). If $n\le 3$ the system is under-determined, which leaves $5-n-1$ free variables to be fully dedicated to the success rate maximization, while for $n=4$ it completely determines all the physical parameters ${\bf p}_{2n}$, leaving no space for further optimization of the success rate. On the contrary, when $n\ge 5$, the system in step (ii) is over-determined and has no exact solutions, so that the $\UG(2n)$ is not able to generate the optimal squeezed core state~(\ref{eq:Core}) that maximizes fidelity with the target state, but may only reach a suboptimal solution. To improve its performance in such scenario, it is necessary to increase the width of the quantum circuit of the scheme, namely to increase the number of modes of the initial Gaussian state, therefore introducing further local squeezings and mode-mixing beam splitters to be suitably optimized.

\begin{figure}
\includegraphics[width=0.95\columnwidth]{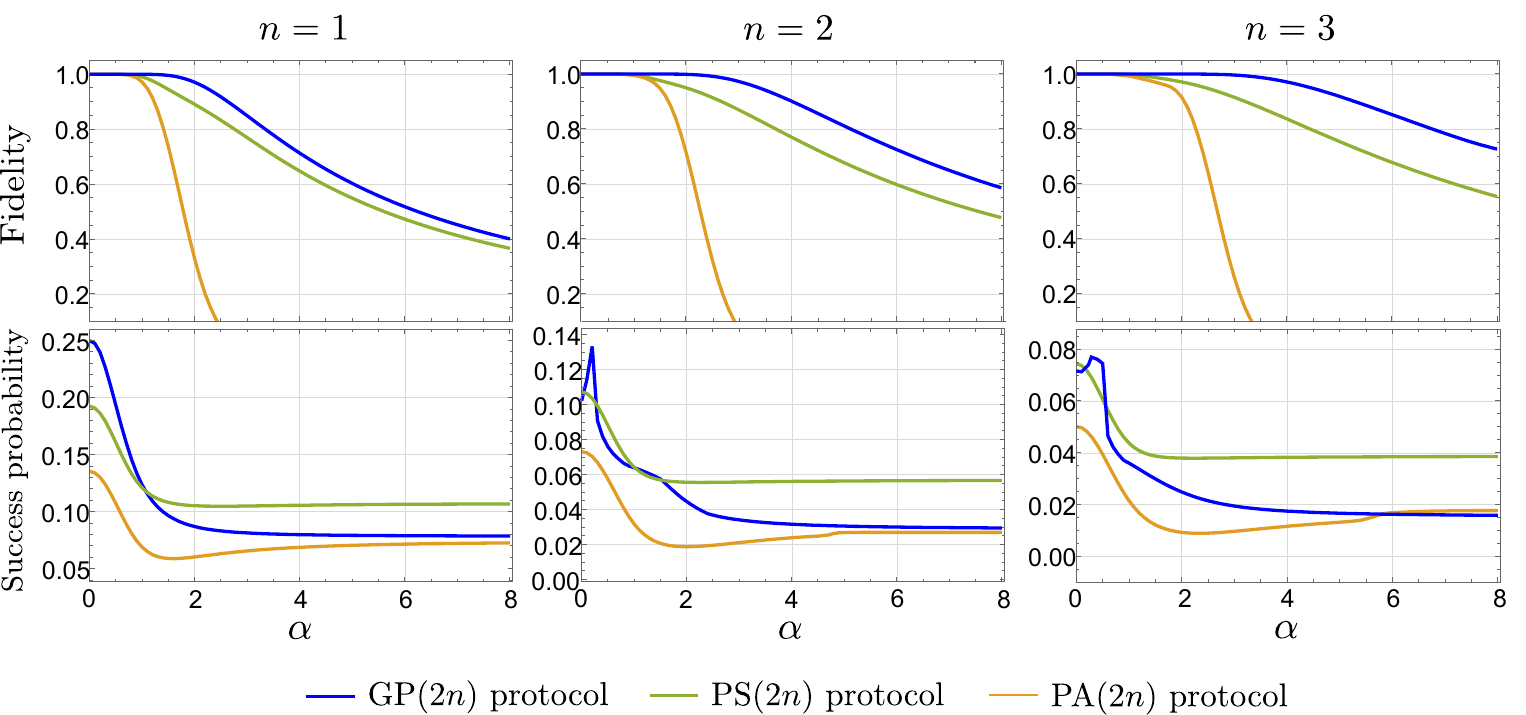} 
\caption{Fidelities $F_{\rm s}^{(2n)}$ (top) and success probability $P_{\rm s}^{(2n)}$ (bottom), $\rm s= \UG, PS, PA$, for $\UG(2n)$, photon-subtraction and addition protocols, respectively, as a function of the target cat amplitude $\alpha$. Lines for success probability for $n\ge2$ look irregular for small $\alpha$, since the adopted optimization algorithm is time consuming, so that the optimized values ${\bf p}_{2n}$ from which $P_{\UG}^{(2n)}$ is computed suffer from numerical errors that affect the smoothness of the curves. In turn, in the limit $\alpha \lesssim 0.5$ the plotted success rates might not be reliably optimal, and slightly larger values of $P_{\UG}^{(2n)}$ might be obtained if a reasonable ansatz on the parameters is available, following the approach outlined in \cite{Jaromir1,Jaromir2}. Still, the plotted rates provide a reasonable indication of the actual orders of magnitude of the $\UG(2n)$ success probability. 
}\label{fig:S02-PlotsGaussianSchemes}
\end{figure}
\begin{figure}
\includegraphics[width=0.45\columnwidth]{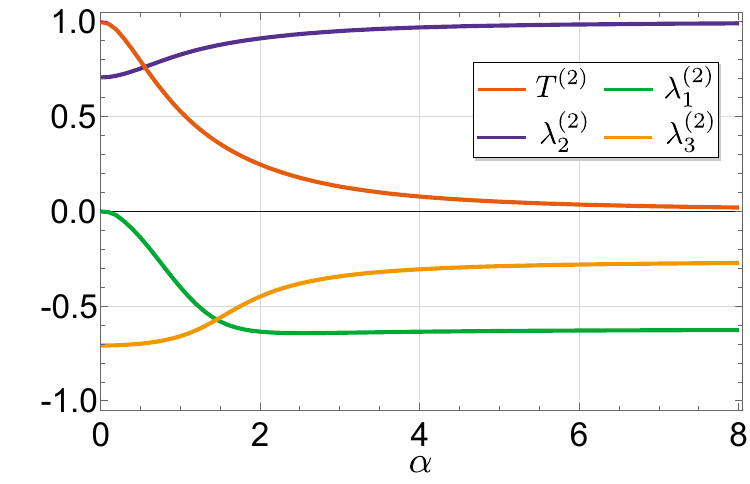}
\caption{Optimized parameters $T^{(2)}$ and $\lambda_j^{(2)}=\tanh(r_j^{(2)})$, $j=1,2,3$, for the $\UG(2)$ protocol. The optimal displacement in this case is null, $\beta^{(2)}=0$. Analogous results are obtained also for cases $n=2,3$, where the resulting setup parameters are piecewise functions of $\alpha$, that suffer from numerical errors due to the time-consuming nature of the adopted optimization algorithm. In those cases, the optimal displacements are nonzero: we have $\beta^{(4)}>0$ for $n=2$, while, for $n=3$, $\beta^{(6)}$ gets purely imaginary values for large $\alpha$, i.e. $(\beta^{(6)})^2<0$.}\label{fig:S03-OptParamtersGP2}
\end{figure}

For comparison, we also consider the performance of the conventional $2n$-photon subtraction [PS($2n$)] and $2n$-photon addition [PA($2n$)] protocols depicted in Fig.~\ref{fig:S01-GP2nSchemes}(b). We note that both these schemes can be re-derived from the $\UG(2n)$ setup by the choices ${\bf p}_{\rm PS}=(r_1,0,0,T,0)$ and ${\bf p}_{\rm PA}=(r_1,-r_1,0, 1/2,\beta)$ and subsequent optimization over the subsets $(r_1,T)$ and $(r_1,\beta)$, respectively. However, for the subtraction case, we would retrieve a variation of the PS($2n$) scheme employing two-mode multiplexing PNR detection, that comes at the cost of a lower success rate, reduced by factor $(2n)!/2^{2n}(n!)^2$ with respect to the single-PNRD scheme displayed in Fig.~\ref{fig:S01-GP2nSchemes}(b). Accordingly, we obtain fidelity of the PS($2n$) and PA($2n$) protocols as $F_{\rm s}^{(2n)}= \max_{{\bf p}_{\rm s}} | \langle \cat^{(+)} |\phi_{2n}^{(+)} ({\bf p}_{\rm s})\rangle|^2$, $\rm s= PS,PA$, with success rates
\begin{align}
P_{\rm PS}^{(2n)}= \frac{2^{2n}(n!)^2}{(2n)!} P_{2n} \left({\bf p}_{\rm PS} \right) \qquad \mbox{and} \qquad P_{\rm PA}^{(2n)}=  P_{2n} \left({\bf p}_{\rm PA} \right) \, ,
\end{align}
to be computed for the optimized values of parameters ${\bf p}_{\rm s}$, respectively.

Plots of $F_{\rm s}^{(2n)}$ and $P_{\rm s}^{(2n)}$, $\rm s= \UG, PS,PA$, for $n=1,2,3$ are reported in Fig.~\ref{fig:S02-PlotsGaussianSchemes}.
As expected, the $\UG(2n)$ protocol outperforms conventional schemes in terms of fidelity, while the success probability is larger than both $P_{\rm PS}^{(2n)}$ and $P_{\rm PA}^{(2n)}$ only for small $\alpha$. Instead, in the limit $\alpha\gg1$, we have $P_{\UG}^{(2)}\approx 7.9\%$, while $P_{\rm PS}^{(2)}\approx 11\%$ and $P_{\rm PA}^{(2)}\approx 7\%$ for case $n=1$; $P_{\UG}^{(4)}\approx 3.0\%$, $P_{\rm PS}^{(4)}\approx 5.6\%$, $P_{\rm PA}^{(4)}\approx 2.7\%$ for $n=2$; and $P_{\UG}^{(6)}\approx 1.6\%$, $P_{\rm PS}^{(6)}\approx 3.9\%$, $P_{\rm PA}^{(6)}\approx 1.8\%$ for $n=3$. 
As a remark we note that, whilst fidelity lines in Fig.~\ref{fig:S02-PlotsGaussianSchemes} are smooth, the lines for success probability look irregular for $n=2,3$ and $\alpha\lesssim 0.5$. This is due to the fact that the fidelity $F_\UG^{(2n)}$ in this regime is close to $1$ and weakly dependent on parameters ${\bf p}$, whereas the success probability $P_{2n} \left({\bf p} \right)$, to be maximized in step (iii) is much more sensitive to small variations of $\bf p$. In turn, the joint optimization procedure outlined above becomes time consuming for $\alpha\ll1$ and the optimized values ${\bf p}_{2n}$ suffer from numerical errors and might not represent global maxima of the function $P_{2n} \left({\bf p} \right)$. 
Nevertheless, the values displayed in Fig.~\ref{fig:S02-PlotsGaussianSchemes} provide a reasonable indication of the actual orders of magnitude of the $\UG(2n)$ success probability. Slightly larger values of $P_{\UG}^{(2n)}$ for $\alpha\ll 1$ could be obtained if a reasonable ansatz on the parameters is available, following the approach outlined in \cite{Jaromir1,Jaromir2}.

For completeness, Fig.~\ref{fig:S03-OptParamtersGP2} reports the optimized parameters ${\bf p}_{2}=(r_{1}^{(2)}, r_{2}^{(2)}, r_{3}^{(2)}, T^{(2)}, \beta^{(2)})$ of the $\UG(2)$ scheme as a function of the target cat amplitude $\alpha$. As we can see, the transmissivity $T^{(2)}$ and the squeezing $\lambda_2^{(2)}=\tanh(r_{2}^{(2)})$ are decreasing and increasing function of $\alpha$, approaching $0$ and $1$ for $\alpha\gg 1$, respectively. This implies that a large squeezing level is needed to achieve large cat generation. Instead, the optimized $r_{1(3)}^{(2)}$ maintain finite values, whereas the displacement $\beta^{(2)}$ is identically null for $n=1$. Instead, the other case studies $n=2,3$ shows nonzero optimal displacements: we have $\beta^{(4)}>0$ for $n=2$, while, for $n=3$, $\beta^{(6)}$ gets purely imaginary values for large $\alpha$, i.e. $(\beta^{(6)})^2<0$.
Moreover, for the basic case $n=1$, we also tried to employ feedforward PNRD to enhance the protocol performance, being a well-known strategy in the context of quantum discrimination \cite{Dolinar1973, Sych2016, Notarnicola2023_HYNORE, Notarnicola2023_HFFRE, Notarnicola_HY_PhN} and quantum communications \cite{Izumi2013, Notarnicola2023_KOR}, but only observed an increase of success probability for $\alpha\ll 1$, and no improvement in the quality of the output state.

%--------------------------------------------------------------------------------------------------------------------------------------------------%
\section{Probabilistic version of the feedforward dispersive protocol}\label{sec:DISP}

\begin{figure}[t]
\includegraphics[width=0.95\columnwidth]{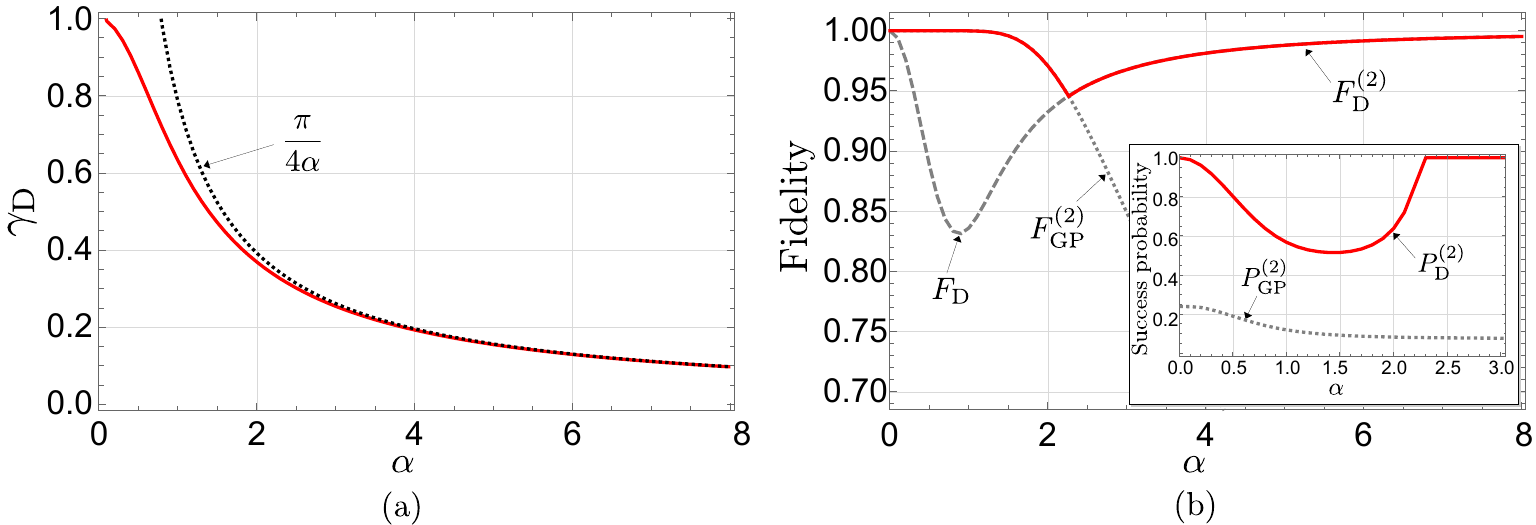}
\caption{(a) Optimized displacement amplitude $\gamma_\DISP$ of the deterministic dispersive protocol as a function of the target cat amplitude $\alpha$: for $\alpha\gg1$ we have $\gamma_\DISP \approx \pi/4\alpha$. (b) Fidelity $F_\DISP^{(2)}$ of the probabilistic dispersive protocol as a function of $\alpha$, compared to $F_{\UG}^{(2)}$ and $F_{\DISP}$, associated with the $\UG$ and deterministic dispersive scheme, respectively. In the inset, success probability $P_{\DISP}^{(2)}$ as a function of $\alpha$, compared to the success rate $P_{\UG}^{(2)}$ of the $\UG(2)$ protocol. %presented in Sec.~\ref{sec:FFversion}. 
We have $P_{\DISP}^{(2)}>P_{\UG}^{(2)}$ and $F_{\DISP}^{(2)}\ge F_{\UG}^{(2)}$, thus demonstrating the dispersive scheme to outperform Gaussian-PNRD protocols for all $\alpha$.}\label{fig:S05-Probabilistic}
\end{figure}

The dispersive coupling protocol presented in the main text, after conditional preparation of the states $|\cat^{(\pm)}\rangle$, involves a displacement operation $D(\pm i\gamma)$ to be performed ever since qubit is measured in the ``-" state, and whose sign is randomly decided with $50\%$ probability, eventually leading to the mixture:
\begin{align}\label{eq:rhoD}
\varrho_\DISP(\gamma) =  p_+ |\cat ^{(+)}\rangle \langle \cat ^{(+)}| + p_- \left( \frac{|\psi'(\gamma)\rangle \langle\psi'(\gamma)| +|\psi'(-\gamma)\rangle \langle\psi'(-\gamma)|}{2} \right) \, ,
\end{align}
with $|\psi'(\gamma)\rangle=D(i \gamma)|\cat ^{(-)}\rangle$ and $p_\pm= (1\pm e^{-2\alpha^2})/2$, that makes the protocol deterministic at the cost of losing the parity of the state, as $|\psi'(\gamma)\rangle$ has nonzero overlap with both the even and odd cat states. Then, the optimal displacement amplitude $\gamma_\DISP$ is chosen to maximize fidelity, namely:
\begin{align}\label{eq:EvenCatFide}
F_\DISP= \max_{\gamma} \, \left\langle \cat^{(+)} \right|\varrho_\DISP(\gamma) \left|  \cat^{(+)} \right\rangle = p_+ +p_- \left[1+ \coth(2\alpha^2)\right] \max_\gamma \left[ \frac{e^{-\gamma^2}}{2} \sin^2(2\alpha \gamma) \right] \, . 
\end{align}
In the large amplitude limit, the optimized displacement amplitude, depicted in Fig.~\ref{fig:S05-Probabilistic}(a), becomes $\gamma_\DISP \approx \pi/4\alpha$, leading to the asymptotic expansion $F_\DISP \approx 1 - \pi^2/32 \alpha^2$, so that for $\alpha \gg 1$ $F_\DISP$ outperforms Gaussian-PNRD schemes.

Now, we also note that the proposed protocol can be further boosted in the low-energy regime to let it outperform the $\UG(2n)$ scheme. This can be done by designing a simple modification of the setup in Fig.~1(a) of the main text, that makes the protocol probabilistic. That is, every time we obtain the ``-" outcome from qubit detection, we perform a random operation: either we apply the conditional displacement $D(\pm i\gamma)$ with probability $q\le 1$, or we discard the state with probability $1-q$. In turn, the resulting protocol produces the average state:
\begin{align}
\varrho_\DISP(q;\gamma) = \frac{1}{ p_+ +q \, p_-} \Bigg[ p_+ |\cat ^{(+)}\rangle \langle \cat ^{(+)}| + \frac{q \, p_-}{2} \Big( |\psi'(\gamma)\rangle \langle\psi'(\gamma)| +|\psi'(-\gamma)\rangle \langle\psi'(-\gamma)| \Big) \Bigg] \, 
\end{align}
with fidelity $F_\DISP(q;\gamma)= \langle \cat^{(+)} |\varrho_\DISP(q;\gamma)|  \cat^{(+)} \rangle$. Now, we perform constrained optimization of the parameters, imposing $F_\DISP(q;\gamma)$ to be larger than $F_\UG^{(2n)}$ of the $\UG(2n)$ protocol, obtaining the $n$-dependent fidelity:
\begin{align}
F_\DISP^{(2n)} = \max_{\substack {q,\gamma \text{ s.t.} \\  F_\DISP(q;\gamma) \ge F_\UG^{(2n)} }  } F_\DISP(q;\gamma)  \, ,
\end{align}
with the corresponding success probability $P_\DISP^{(2n)}= p_+ +q \, p_-$ to be computed for the resulting optimized value of $q$. Plots of $F_\DISP^{(2n)}$ and $P_\DISP^{(2n)}$ for the basic case $n=1$ are depicted in Fig.~\ref{fig:S05-Probabilistic}(b), compared to both the $\UG(2)$ and the deterministic dispersive schemes. We observe that for $\alpha \lesssim 2.3$, we have $F_\DISP^{(2)} = F_\UG^{(2)}$, proving dispersive coupling to reproduce the results of Gaussian method albeit with larger success rate, as $P_{\DISP}^{(2)}>P_{\UG}^{(2)}$. Instead, for larger $\alpha$, we have $F_\DISP^{(2)} = F_\DISP$ and the success probability goes to $1$, retrieving the standard deterministic scheme previously introduced. We conclude that the probabilistic version of the dispersive protocol genuinely beats Gaussian coupling for all $\alpha$. However, as we are mostly interested in large cats generation, in the main text we always considered the protocol in its deterministic version.

%--------------------------------------------------------------------------------------------------------------------------------------------------%
\section{Comparison between dispersive and $\UG(2n)$ protocols}\label{sec:COMP}

While in the main text we provide comparison between Gaussian-PNRD and dispersive protocols in terms of fidelity and Wigner function interference, in this Section, we extend the analysis to other relevant aspects, namely robustness of the produced states with respect to additional losses, homodyne distribution and distillable squeezing.
%------------%
\subsection{Robustness to additional losses}
First of all, we investigate robustness with respect to subsequent optical losses of the output states $|\phi_{2n}^{(+)}\rangle=|\phi_{2n}^{(+)}({\bf p}_{2n})\rangle$ and $\rho_\DISP=\rho_\DISP(\gamma_\DISP)$ produced by the $\UG(2n)$ and dispersive protocols (in its deterministic version), computed for the optimized parameters ${\bf p}_{2n}$ and $\gamma_\DISP$ derived in the previous Sections. To this aim, we model losses in terms of a beam splitter with transmissitivity $\tau \le 1$, and, in particular, we address the relevant regime $1-\tau \ll 1$. 

For the $\UG$ case, the state $|\phi_{2n}^{(+)}\rangle= \sum_{m=0}^\infty c_m |m\rangle$, after passage through the beam splitter, decoheres into:
\begin{align}
\rho_{2n}(\tau)= \Tr_{l} \left[ U_{\rm BS}(\tau) |\phi_{2n}^{(+)}\rangle |0\rangle_l \right] = \sum_{j,k=0}^\infty \, \sum_{m=j}^\infty c_m c^*_{m-j+k} (1-\tau)^{m-j} \sqrt{\binom{m}{j} \binom{m-j+k}{k} \tau^{j+k} } |j\rangle \langle k| \, ,
\end{align}
where $U_{\rm BS}(\tau)$ is the beam splitter unitary operator and mode $l$ refers to the second-port mode entering the splitter. Eventually, the mixed state $\rho_{2n}(\tau)$ loses even parity and gets fidelity with the original target state equal to $F_\UG^{(2n)}(\tau)= \langle \cat^{(+)}|\rho_{2n}(\tau)|\cat^{(+)}\rangle$.
On the contrary, in the dispersive protocol, the states $|\cat^{(+)}\rangle$ and $|\psi'(\pm \gamma_\DISP)\rangle=D( \pm i\gamma_\DISP) |\cat^{(-)}\rangle$ conditionally prepared at the qubit measurement stage, after passage through the loss beam splitter are turned into:
\begin{subequations}\label{eq:losses}
\begin{align}
\rho_+(\tau)&=  \Tr_{l} \left[ U_{\rm BS}(\tau) |\cat^{(+)}\rangle |0\rangle_l \right] \nonumber \\
&= \frac{1}{{\cal N}_+^2} \left\{ |\sqrt{\tau}\alpha\rangle\langle \sqrt{\tau}\alpha| + |-\sqrt{\tau}\alpha\rangle\langle- \sqrt{\tau}\alpha| + e^{-2(1-\tau)\alpha^2} \left(  |\sqrt{\tau}\alpha\rangle\langle - \sqrt{\tau}\alpha| + {\rm h.c.} \right) \right\} \, ,  \label{eq:rho+} \\[2ex]
\rho'(\tau;\gamma_\DISP)&=  \Tr_{l} \left[ U_{\rm BS}(\tau) |\psi'(\gamma_\DISP)\rangle |0\rangle_l \right] \nonumber \\
&= \frac{1}{{\cal N}_-^2} \Big\{ |\sqrt{\tau}(\alpha+i\gamma_\DISP)\rangle\langle \sqrt{\tau}(\alpha+i\gamma_\DISP)| + |\sqrt{\tau}(-\alpha+i\gamma_\DISP)\rangle\langle \sqrt{\tau}(-\alpha+i\gamma_\DISP)| \nonumber \\
& \hspace{2.cm}- e^{-2(1-\tau)\alpha^2} \left(  e^{2 i \tau \alpha \gamma_\DISP} |\sqrt{\tau}(\alpha+i\gamma_\DISP)\rangle\langle \sqrt{\tau}(-\alpha+i\gamma_\DISP)| + {\rm h.c.} \right) \Big\} \, ,
\end{align}
\end{subequations}
with ${\cal N}_\pm= \sqrt{2(1\pm \exp(-2\alpha^2))}$. The average output mixture then becomes:
\begin{align}
\varrho_\DISP(\tau) =  p_+\rho_+(\tau) + \frac{p_-}{2} \Big[\rho'(\tau;\gamma_\DISP) +\rho'(\tau;-\gamma_\DISP)\Big] \, ,
\end{align}
with fidelity $F_\DISP(\tau)= \langle \cat^{(+)}|\varrho_\DISP(\tau )|\cat^{(+)}\rangle$.
%, to be computed for the optimized value $\gamma_\DISP$ of the ideal dispersive protocol.

Plots of $F_\UG^{(2n)}(\tau)$ and $F_\DISP(\tau)$ are depicted in Fig.~2 of the main text for $\tau=0.999$. As we can see from Eq.s~(\ref{eq:losses}), losses have the twofold consequence of both amplitude reduction of the resulting cat-like state and exponential suppression of coherences of $\varrho_\DISP(\tau)$. This latter effect makes $F_\DISP(\tau)$ lose the asymptotic behavior of the ideal state preparation case even in the presence of miniscule losses, being turned into the decreasing function:
\begin{align}
F_\DISP(\tau) \approx e^{-\alpha^2 \left(1-\sqrt{\tau}\right)^2} \frac{1+e^{-2\alpha^2(1-\tau)}}{2} \left( 1 - \frac{\pi^2}{32 \alpha^2} \right) \qquad \mbox{for }\alpha \gg 1 \,.
\end{align}
Analogous results are also obtained for the target cat $|\cat^{(+)}\rangle$, being turned after the losses into the state $\rho_+(\tau)$ in Eq.~(\ref{eq:rho+}), with corresponding fidelity:
\begin{align}
F_+(\tau) \approx e^{-\alpha^2 \left(1-\sqrt{\tau}\right)^2} \frac{1+e^{-2\alpha^2(1-\tau)}}{2} \qquad \mbox{for }\alpha \gg 1 \,,
\end{align}
reported as a gray line in Fig.~2 of the main text.
%by considering fidelity with respect to a target cat state with amplitude $\sqrt{\tau}\alpha$, as deviations from the lossless case in the limit $\tau\sim 1$ is dominantly due to coherence reduction, while the energy loss effect only weakly affect the quality of the resulting states.

\begin{figure*}[t]
\includegraphics[width=0.65\linewidth]{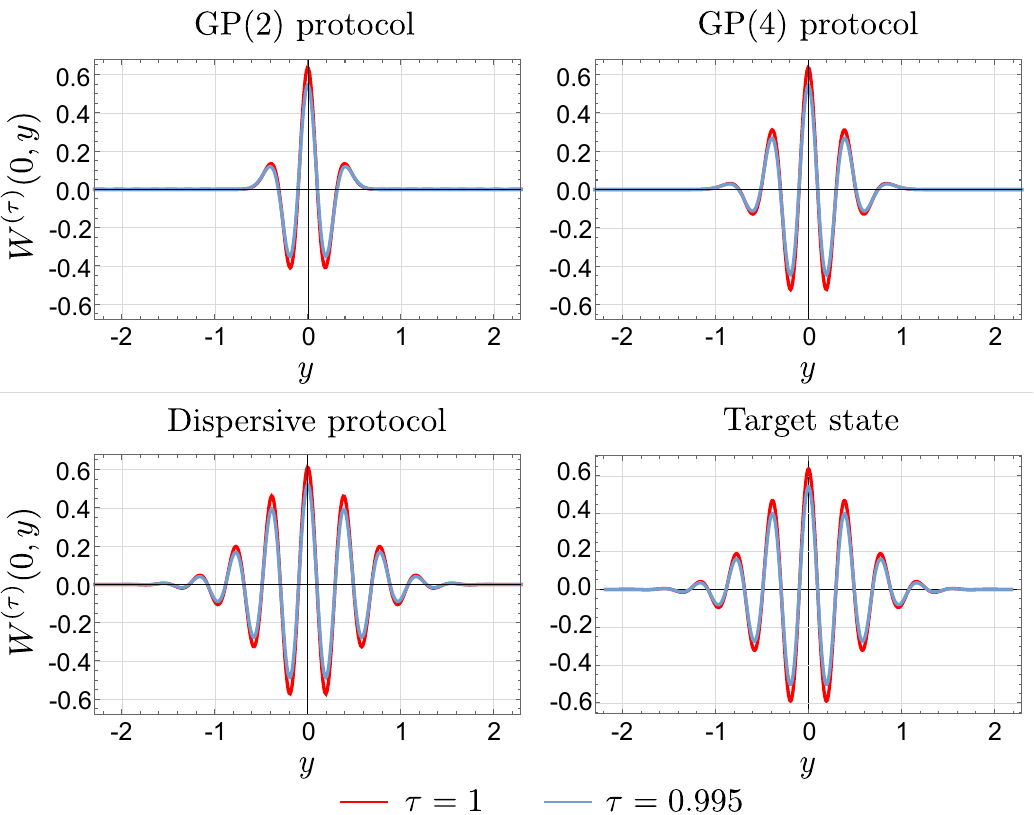}
\caption{Oscillations of the Wigner function $W^{(\tau)}(0,y)$ along the $y$ axis for the $\UG(2n)$ protocol, with $n=1,2$, and the dispersive protocol for $\alpha=4$ in the presence of optical losses of transmissivity $\tau\le 1$.}\label{fig:S06-WignerLoss}
\end{figure*}

Furthermore, we assess impact of the losses on phase space properties, by computing the Wigner function \cite{Olivares2021, NotarnicolaTesi}
\begin{align}\label{eq:Wigner}
W^{(\tau)}(x,y)= \frac{2}{\pi} \sum_{n=0}^{\infty} (-1)^n \langle n | D^\dagger(\zeta) \, \rho \, D(\zeta) |n\rangle \,, \qquad \rho=\rho_{2n}(\tau),\varrho_\DISP(\tau),\varrho_+(\tau) \, ,
\end{align}
%of both states $\rho_{2n}(\tau)$, $\varrho_\DISP(\tau)$ and $\varrho_+(\tau)$, 
where $D(\zeta)$, $\zeta=x+i y$, is a displacement operation and $\{|n\rangle\}_n$ is the Fock basis, depicted in Fig.~\ref{fig:S06-WignerLoss}. For all the considered states, losses induce a $\tau$-sensitive reduction of Wigner negativity, being consequence of loss of coherences, with progressive suppression of the interference fringes. In particular, the minimum value of the dispersive Wigner function reads: 
\begin{align}
W_\DISP^{(\tau)}(0,\bar{y})  &\approx W_{\rm cat}^{(+)} (0,\bar{y}) +  \frac{1}{\pi} \left[ 2-e^{-2 \alpha ^2 (1-\tau )} \left(\frac{\cos \left(\pi  \left(\tau -\sqrt{\tau }\right)\right)+\cos \left(\pi  \left(\tau +\sqrt{\tau }\right)\right)}{2}-\cos \left(\pi  \sqrt{\tau }\right)\right) \right] + \frac{\pi }{8\alpha^2}  \Bigg[-2 + e^{-2 \alpha ^2 (1-\tau )} \times \nonumber \\[1ex]
&\left(\frac{\left(1-\sqrt{\tau }\right)^2 \cos \left(\pi  \left(\tau -\sqrt{\tau }\right)\right)+\left(1+\sqrt{\tau }\right)^2 \cos \left(\pi  \left(\tau +\sqrt{\tau }\right)\right)}{2}-\cos \left(\pi  \sqrt{\tau }\right)\right)\Bigg]  \qquad \mbox{for }\alpha \gg 1  \,,
\end{align}
where
$$W_{\rm cat}^{(+)}(x,y) =\frac{1}{\pi \left(1+e^{-2 \alpha ^2}\right)} \Bigg[ 2 e^{-2 \left(x^2+y^2\right)} \cos (4 \alpha  \,  y)+ e^{-2 \left((x-\alpha )^2 +y^2 \right)}+e^{-2 \left((x+\alpha)^2+y^2\right)} \Bigg]
%W_{\rm cat}^{(+)}(x,y) = [ 2 e^{-2 (x^2+y^2)} \cos (4 \alpha  \,  y)+ e^{-2 ((x-\alpha )^2 +y^2)}+e^{-2 ((x+\alpha)^2+y^2)}]/[\pi (1+e^{-2 \alpha ^2})]
$$
is the target state Wigner function and $\bar{y}$ being the point where both $W_\DISP^{(\tau)}$ and $W_{\rm cat}^{(+)}$ gets minimum. In the limit $\tau \sim 1$ we simplify the previous expression as:
\begin{align}\label{eq:WminLOSS}
W_\DISP^{(\tau)}(0,\bar{y})  \approx W_{\rm cat}^{(+)} (0,\bar{y}) +  \frac{1}{\pi} \left(1-e^{-2\alpha^2(1-\tau)}\right)  + \frac{\pi }{8\alpha^2} \Bigg[-2+ (2+\tau) e^{-2\alpha^2(1-\tau)}\Bigg]  \qquad \mbox{for }\alpha \gg 1 \mbox{  and  } 1-\tau \ll 1 \,.
\end{align}
Similarly, for the decohered target state $\rho_+(\tau)$ we get:
\begin{align}\label{eq:WtgtLOSS}
W_+^{(\tau)}(0,\bar{y})  \approx W_{\rm cat}^{(+)} (0,\bar{y}) +  \frac{1}{\pi} \left(1-e^{-2\alpha^2(1-\tau)}\right) \qquad \mbox{for }\alpha \gg 1 \mbox{  and  } 1-\tau \ll 1 \,,
\end{align}
determining exponential suppression of Wigner negativity, proportional to $1-e^{-2\alpha^2(1-\tau)}$. 
On the contrary, in the lossless case $\tau=1$, Eq.~(\ref{eq:WminLOSS}) reduces to the ideal case scenario discussed in the main text, namely $W_\DISP(0,\bar{y})  \approx W_{\rm cat}^{(+)} (0,\bar{y}) +\pi/8\alpha^2$, while Eq.~(\ref{eq:WtgtLOSS}) re-approaches the target value $W_{\rm cat}^{(+)} (0,\bar{y})$.

%------------%
\subsection{Distillable squeezing}

\begin{figure*}[t]
\includegraphics[width=0.95\linewidth]{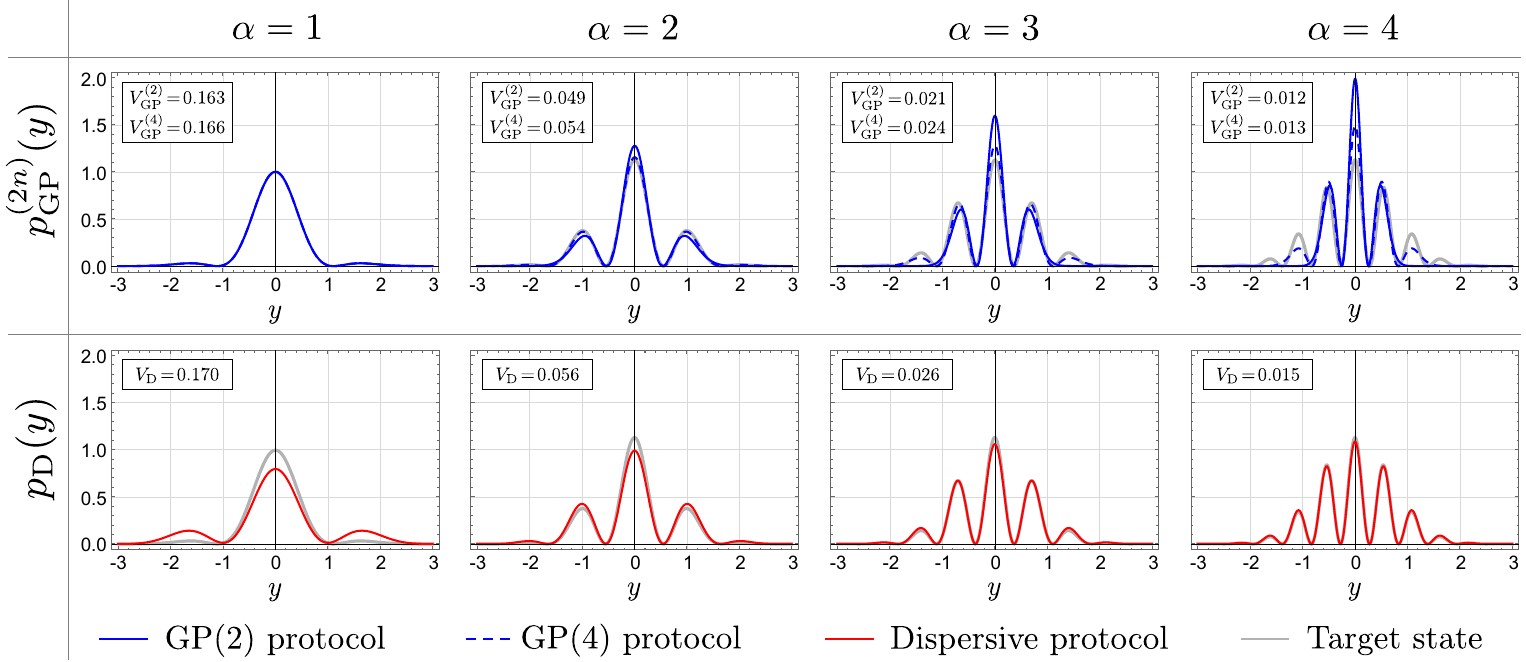}
\caption{Homodyne momentum-like distribution $p(y)$ for the $\UG(2n)$ protocol, with $n=1,2$, (top row) and the dispersive protocol (bottom row) for different values of the cat amplitude $\alpha$. The boxes report corresponding values of distillable squeezing variance, plotted in Fig.~\ref{fig:S08-Distillable}.}\label{fig:S07-Homodyne}
\end{figure*}

For a further analysis of the phase space quantum interference of both the $\UG(2n)$ and dispersive protocols, it is also interesting to evaluate the momentum-like probability distribution of states $|\phi_{2n}^{(+)}\rangle$ and $\rho_\DISP$, implemented by homodyne measurement of quadrature $\hat{y}=i (a\dag-a)/\sqrt{2}$, expressed in canonical units \cite{NotarnicolaTesi}. The corresponding probability distribution is obtained by projection over the momentum eigenstates, namely:
\begin{align}\label{eq:HomodyneP}
p(y) =  \langle y |\rho |y\rangle \nonumber= \frac{e^{-y^2}}{\sqrt{\pi}} \sum_{n,m=0}^{\infty} \langle n | \rho | m \rangle \, \, \frac{H_n(y) H_m(y)}{\sqrt{2^{n+m} n!m!}} \, \, e^{i \frac{\pi}{2} (n-m)} \,,
\end{align}
where $|y\rangle=e^{-y^2/2}/\pi^{1/4} \sum_{n} e^{-i \pi n/2} H_n(y)/\sqrt{2^n n!}  \, |n\rangle$ is the eigenstate of $\hat{y}$ associated with outcome $y$, and $H_n(y)$ being the $n$-th Hermite polynomial \cite{Olivares2021, Notarnicola2024_Kerr}.
The corresponding distributions for the protocols under consideration, that is $p_\UG^{(2n)}(y) =  |\langle y |\phi_{2n}^{(+)}\rangle|^2$ and $p_\DISP(y) =  \langle y |\rho_\DISP |y\rangle$
%, to be computed for the optimized displacement $\gamma_\DISP$, 
are reported in Fig.~\ref{fig:S07-Homodyne} for different values of the target cat amplitude $\alpha$. Given the finite stellar rank, the homodyne probability of the $\UG(2n)$ scheme exhibits only $2n+1$ maxima, with the peak in the origin higher than the target state, for which the homodyne probability reads 
$$p_{\rm cat}^{(+)} (y)= \frac{e^{-y^2+ 2 \alpha ^2}}{\sqrt{\pi}(1+e^{2 \alpha ^2})} \left(1+ \cos(2 \sqrt{2} \alpha  \, y)\right) \, .$$
As for the Wigner function, there is accordance with the target state only in the kitten regime $\alpha\ll 1$, whereas for $\alpha \gtrsim 2$ the shapes of the two distributions become appreciably different. 
The scenario is reversed for the dispersive protocol, in which case we observe infinitely many oscillations of $p_\DISP(y)$ for any $\alpha$, and a central peak in the origin lower than the target state, closing the gap between them for increasing $\alpha$.

\begin{figure}[t]
\includegraphics[width=0.45\columnwidth]{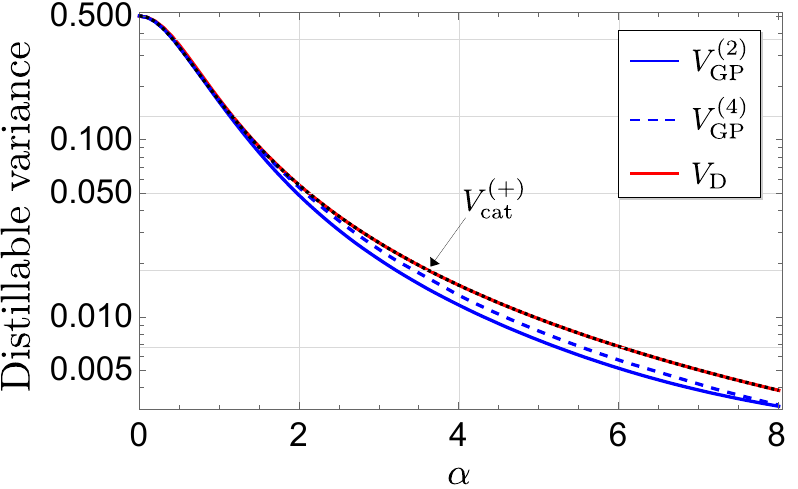}
\caption{Log plot of distillable squeezing variance for both the $\UG(2n)$ protocol, with $n=1,2$, and the dispersive protocol as a function of the target cat amplitude $\alpha$. The dispersive variance $V_\DISP$ is almost equal to the distillable variance $V_{\rm cat}^{(+)}$ of the target cat state.}\label{fig:S08-Distillable}
\end{figure}

To investigate quality of the interference fringes of the previous probabilties, we consider the distillable squeezing variance, equal to:
\begin{align}
V= \frac{p(0)}{|p''(0)|} \, ,
\end{align}
to be computed from the homodyne distribution $p(y)$ around the global maxima $y=0$, where $p''(y)= d^2p(y)/dy^2$ is the second derivative \cite{Distillation}.
It represents the maximum amount of squeezing achievable after a universal distillation scheme by beam splitters and quadrature measurements, in the asymptotic limit of infinitely many copies of the input state, therefore it provides a witness of nonclassicality whenever its value is below vacuum fluctuations, i.e. $V<1/2$ \cite{Hage2007,Marek2007,Distillation,DarrenDSQ, Fiurasek2025}.
Fig.~\ref{fig:S08-Distillable} reports the distillable variances $V_\UG^{(2n)}$ and $V_\DISP$ for both the $\UG(2n)$ and dispersive protocols as a function of $\alpha$, compared to the target state for which:
\begin{align}
V_{\rm cat}^{(+)}= \frac{1}{2(1+2\alpha^2)} \, ,
\end{align}
that for $\alpha\gg1$ approaches $V_{\rm cat}^{(+)} \approx 1/4\alpha^2$. As we can see from the plot, we have $V<1/2$ for all $\alpha$ for both the $\UG$ and dispersive schemes. The $\UG$ state has lower variance than the target state, as its homodyne distribution is more peaked around the origin due to its finite stellar rank, whereas the dispersive variance $V_\DISP$ is almost equal to the distillable variance of the target cat state, and for $\alpha \gg 1$ we have:
\begin{align}
V_\DISP \approx  \frac{1}{4\alpha^2} - \frac{1}{8\alpha^4}\approx V_{\rm cat}^{(+)} - \frac{1}{8\alpha^4} \, .
\end{align}

%--------------------------------------------------------------------------------------------------------------------------------------------------%
\section{Robustness of dispersive protocol to experimental imperfections }\label{sec:COMP}

%------------%
\subsection{Imperfect realization of dispersive coupling: derivation of the model}
\begin{figure*}[t]
\includegraphics[width=0.9\linewidth]{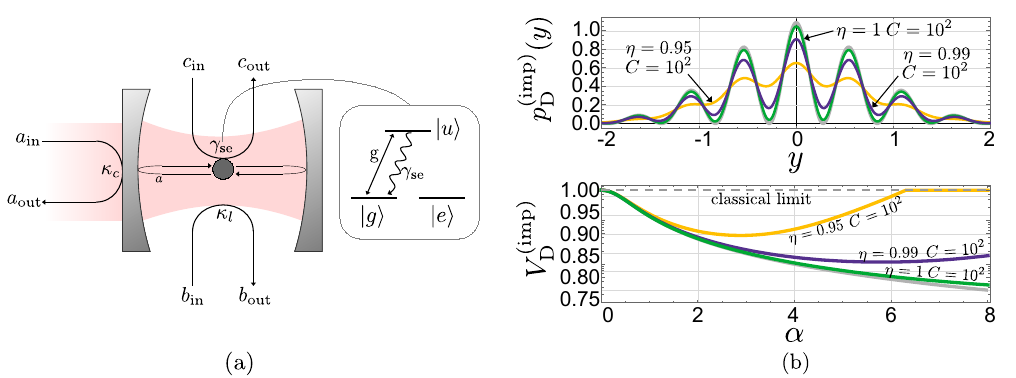}
\caption{(a) Modelization of the dispersive coupling cavity QED setup in terms of the input-output formalism. (b) Homodyne momentum-like distribution $p_\DISP^{(\IMP)}(y)$ for $\alpha=4$ and distillable squeezing variance $V_\DISP^{(\IMP)}$ of the imperfect dispersive coupling protocol for different values of $C$ and $\eta$. The gray line corresponds to the ideal dispersive scheme.}\label{fig:S09-Imperfect}
\end{figure*}

As explained in the main text, the dispersive interaction needed for the proposed protocol is effectively implemented by the reflection of an optical beam to a $\Lambda$-type atom cavity schematized in Fig.~\ref{fig:S09-Imperfect}(a). In more detail, the cavity mode $a$ is coupled to impinging free-space field with rate $\kappa_c\ge0$ and to unwanted scattering and loss into other spatial modes with rate $\kappa_l\ge0$. The $\Lambda$-type atom is composed of two degenerate levels $|g\rangle$ and $|e\rangle$ and an excited level $|u\rangle$, where only  transition $|g\rangle \to |u\rangle$ is resonantly coupled to the cavity field by Jaynes-Cummings interaction with gain $\rm g\ge0$. The transition $|g\rangle \to |u\rangle$ is further associated with spontaneously decay rate $\gamma_{\rm se}\ge0$ \cite{Rempe2019, Teja2023, Ulrik2022, LachmannQUANTUM}.

In this Section, we provide theoretical description for the present system within the framework of the input-output formalism, considering the realistic case of finite cooperativity  $C = g^2/(\kappa\gamma_{\rm se}) <\infty$, $\kappa=\kappa_c+\kappa_l$, and non-unit escape efficiency $\eta=\kappa_c/\kappa <1$.
Without loss of generality, we first solve the problem in the subspace of the atom Hilbert space spanned by $\{|g\rangle, |u\rangle\}$, and then address the $\{|e\rangle, |u\rangle\}$ subspace. The atom-cavity Hamiltonian reads: 
\begin{align}
H_0= \omega \, a^\dagger a + \omega \, |u\rangle\langle u | + {\rm g} \left(\bmsigma_{gu} \,a^\dagger+\bmsigma_{ug} \,a \right) \, ,
\end{align}
where both the cavity and atom are resonant at frequency $\omega>0$, while $\bmsigma_{gu}=|g\rangle\langle u|$ and  $\bmsigma_{ug}=|u\rangle\langle g|$ are the atomic lowering and raising operators, respectively.
We describe coupling to the environment at time $t$ by three additional temporal modes: $\{a_t\}_t$, related to the impinging free-space mode with $[a_t, a_s]=\delta(t-s)$; $\{b_t\}_t$, describing the further unwanted losses; and $\{c_t\}_t$, accounting for the spontaneous emission radiated field \cite{Breuer2002}. Accordingly, the system-environment interaction Hamiltonian at time $t$ is equal to $H_{\rm int}(t)= i\sqrt{\kappa_c} (a_t a^\dagger - a_t^\dagger a) +  i\sqrt{\kappa_l} (b_t a^\dagger - b_t^\dagger a) +i\sqrt{\gamma_{\rm se}} (c_t \bmsigma_{ug} - c_t^\dagger \bmsigma_{gu})$. The temporal modes $\{a_t\}_t$ evolve as $\dot{a}_s= i \, [H_{\rm int}(t), a_s]= -\sqrt{\kappa_c} \delta(t-s) a$, with solution $a_s(t)= a_s(0) - \sqrt{\kappa_c} \Theta(t-s) a$, where $\Theta(x)$ is the Heaviside Theta function, defined by $\Theta(x)=1$ for $x>0$, $\Theta(x)=0$ for $x<0$, and $\Theta(0)=1/2$. 
In turn, we derive the boundary condition \cite{Collett1984, Wiseman2009}:
\begin{align}\label{eq:BC}
a_{\rm out}= a_{\rm in} - \sqrt{\kappa_c} \, a \, ,
\end{align}
where $a_{\rm out}=a_t(t+\Delta t)$, $\Delta t\ll 1$, and $a_{\rm in}=a_t(0)$ are the output (input) mode after (before) the interaction takes place at time $t$. Analogous results are obtained also for modes $\{b_{\rm in}, b_{\rm out}\}$ and $\{c_{\rm in}, c_{\rm out}\}$, namely: $b_{\rm out}=b_{\rm in} - \sqrt{\kappa_l} \, a$ and $c_{\rm out}= c_{\rm in} - \sqrt{\gamma_{\rm se}} \, \bmsigma_{gu}$.

Now, exploiting the boundary conditions, we retrieve the corresponding Langevin equation for any arbitrary operator $A$ of the atom-cavity system as:
\begin{align}
\dot{A}(t)&= i \, [H_0+H_{\rm int}(t), A] \nonumber \\[1ex]
&= i \, [H_0, A] +[a^\dagger, A] \left(\frac{\kappa}{2} a - \sqrt{\kappa_c} a_{\rm in} - \sqrt{\kappa_l} b_{\rm in} \right) - \left(\frac{\kappa}{2} a^\dagger - \sqrt{\kappa_c} a_{\rm in}^\dagger - \sqrt{\kappa_l} b_{\rm in}^\dagger \right) [a, A] \nonumber \\[1ex]
& \hspace{1.cm} + [\bmsigma_{ug}, A] \left(\frac{\gamma_{\rm se}}{2} \bmsigma_{gu} - \sqrt{\gamma_{\rm se}} c_{\rm in} \right) -  \left(\frac{\gamma_{\rm se}}{2} \bmsigma_{ug} - \sqrt{\gamma_{\rm se}} c_{\rm in}^\dagger  \right) [\bmsigma_{gu}, A]\, .
\end{align}
For our purposes it suffices to consider the choices $A=a, \bmsigma_{gu}$, obtaining:
\begin{align}
\dot{a}(t)&=- i \omega a - i {\rm g} \bmsigma_{gu} - \frac{\kappa}{2} a + \sqrt{\kappa_c} a_{\rm in} + \sqrt{\kappa_l} b_{\rm in}  \, , \nonumber \\
\dot{\bmsigma}_{gu}(t)&=- i \omega  \bmsigma_{gu} - \left(\bmsigma_{gg}-\bmsigma_{uu}\right) \left( i {\rm g} a + \frac{\gamma_{\rm se}}{2}  \bmsigma_{gu}- \sqrt{\gamma_{\rm se}} c_{\rm in} \right) \, ,
\end{align}
with $\bmsigma_{gg(uu)}=|g(u)\rangle \langle g(u)|$.
We further assume the atom to be weakly excited, so that $\bmsigma_{ee} \approx 0$ and $\bmsigma_{gg} \approx \mathbbm{1}$, and, for convenience, we move to the reference frame rotating at frequency $\omega$, i.e. perform the transformation $(a,\bmsigma_{gu},a_{\rm in},b_{\rm in},c_{\rm in})\to (a,\bmsigma_{gu},a_{\rm in},b_{\rm in},c_{\rm in}) e^{-i\omega t}$. Eventually, we get the dynamical equations:
\begin{align}
\begin{cases}
\dot{a}(t)= - i {\rm g} \,\bmsigma_{gu} - \frac{\kappa}{2}\, a + \sqrt{\kappa_c}\, a_{\rm in} + \sqrt{\kappa_l}\, b_{\rm in}  \, , \\[1ex]
\dot{\bmsigma}_{gu}(t)= - i {\rm g}\, a - \frac{\gamma_{\rm se}}{2}  \,\bmsigma_{gu}+ \sqrt{\gamma_{\rm se}}\, c_{\rm in} \, .
\end{cases}
\end{align}
Finally, we look for steady state solutions by setting $\dot{a}(t)=0$ and $\dot{\bmsigma}_{gu}(t)=0$, obtaining the final modes transformation:
\begin{align}\label{eq:MODESTRANSF}
    a_{\rm out} =  \left( 1 -\frac{2\eta}{1+4C} \right) \, a_{\rm in} - \frac{2\sqrt{\eta(1-\eta)}}{1+4C} b_{\rm in} +\frac{4i\sqrt{\eta C}}{1+4C} c_{\rm in} \, ,
\end{align}
where we plugged in condition~(\ref{eq:BC}).
The modes transformation for the non-interacting transition $|e\rangle \to |u\rangle$ are also obtained from~(\ref{eq:MODESTRANSF}) by setting $C=0$, such that, overall, the input free-space mode $a_{\rm in}$ is conditonally transformed into:
\begin{align}\label{eq:CondTransformation}
a_{\rm in} \to   \left(\sqrt{\eta_g} \, a_{\rm in} - \sqrt{\eta'-\eta_g} \, b_{\rm in} + i \sqrt{1-\eta'} \, c_{\rm in} \right) \otimes |g\rangle \langle g| - \left(\sqrt{\eta_e} \, a_{\rm in} + \sqrt{1-\eta_e} \, b_{\rm in} \right)\otimes |e\rangle \langle e| \, ,
\end{align}
where:
\begin{subequations}
\begin{align}
\eta_g &= \left(1 - \frac{2\eta}{1+4C}\right)^2\, , \\
\eta' &= 1- \frac{16\eta C}{(1+4C)^2} \, ,\\
\eta_e &= (1 - 2\eta)^2\, , 
\end{align}
\end{subequations}
and assuming $\eta>1/2$ such that $1-2\eta= - \sqrt{\eta_e}<0$. These results justify the effective modeling of imperfect dispersive coupling in terms of conditional loss acting on the optical field, with qubit-state dependent transmissivity $\eta_{g(e)}$, respectively. We also note that, in the limits $C\to \infty$ and $\eta=1$, Eq.~(\ref{eq:CondTransformation}) reduces to $a_{\rm in} \to   a_{\rm in}  \otimes |g\rangle \langle g| - a_{\rm in}\otimes |e\rangle \langle e| = a_{\rm in}  \bmsigma_{\rm z}$, thus implementing the conditional $\pi$ phase-shift rotation corresponding to ideal dispersive coupling.

In light of this, given the input states $|\alpha\rangle$ and $|+\rangle=(|g\rangle+|e\rangle)/\sqrt{2}$ for the optical field and atom, respectively, the global output state after the interaction reads:
\begin{align}
|\Psi\rangle_{\rm out} = \frac{1}{\sqrt{2}} \left\{ |\sqrt{\eta_g}\alpha\rangle_{a_{\rm out}} |-\sqrt{\eta'-\eta_g}\alpha\rangle_{b_{\rm out}} |i \sqrt{1-\eta'} \alpha \rangle_{c_{\rm out}} \otimes |g\rangle + |-\sqrt{\eta_e}\alpha\rangle_{a_{\rm out}} |-\sqrt{1-\eta_e}\alpha\rangle_{b_{\rm out}} |0\rangle_{c_{\rm out}} \otimes |e\rangle \right\} \, .
\end{align}
Then, we perform partial trace over the unwanted modes $b_{\rm out},c_{\rm out}$, and measure the atom in the $\{|\pm\rangle\}$ basis, leaving mode $a_{\rm out}$ in the mixture:
\begin{align}\label{eq:sigmaIMP}
\sigma_\pm^{(\IMP)} = \frac{1}{4 p_{\pm}^{(\IMP)} } \Big\{ |\sqrt{\eta_g} \alpha\rangle\langle \sqrt{\eta_g} \alpha| +  |-\sqrt{\eta_e} \alpha\rangle\langle -\sqrt{\eta_e} \alpha|  \pm e^{-\Gamma \alpha^2/2} \Big(  |\sqrt{\eta_g} \alpha\rangle\langle- \sqrt{\eta_e} \alpha|+|-\sqrt{\eta_e} \alpha\rangle\langle \sqrt{\eta_g} \alpha| \Big) \Big\} \, ,
\end{align}
with $\Gamma=2-\eta_g-\eta_e+2\sqrt{(1-\eta_e)(\eta'-\eta_g)}$, and detection probabilities $p_{\pm}^{(\IMP)}=(1\pm e^{-[\Gamma+ (\sqrt{\eta_g}+\sqrt{\eta_e})^2]\alpha^2/2})/2$. The average output state is turned into:
\begin{align}\label{eq:rhoIMP}
\varrho_\DISP^{(\IMP)}(\gamma) = p_+^{(\IMP)}  \sigma_+^{(\IMP)} +\frac{p_-^{(\IMP)}}{2} \left[ D( i \gamma)  \sigma_-^{(\IMP)}  D\dag( i \gamma) + D(- i \gamma)  \sigma_-^{(\IMP)}  D\dag(- i \gamma) \right] \, ,
\end{align}
for which we compute optimized fidelity   $F_\DISP^{(\IMP)}= \max_{\gamma} \, \langle \cat^{(+)} |\varrho_\DISP^{(\IMP)}(\gamma)|  \cat^{(+)} \rangle$ and Wigner negativity, discussed in the main text.
Here, in Fig.~\ref{fig:S09-Imperfect}(b) we report also the homodyne momentum-like distribution $p_\DISP^{(\IMP)}(y)$ for $\alpha=4$ (top row) and the corresponding distillable squeezing variance  $V_\DISP^{(\IMP)}$ (bottom row) as a function of $\alpha$. As we can see, in the presence of finite cooperativity and non-unit escape efficiency, the homodyne probability exhibits nonzero minima and reduced heights of maxima, such that, for low values of $C$ and $\eta$, they approach
each other and reduce the interference visibility. In turn, the distillable variance increases for large $\alpha$, and, for sufficiently low $C$ and $\eta$, eventually reach the classical limit $V_\DISP^{(\IMP)}=1/2$, corresponding to vacuum fluctuations, for $\alpha \gg 1$.  On the contrary, in the opposite regime $C\gg 1$ and $1-\eta \ll 1$, we find the analytical expansion:
\begin{align}
V_\DISP^{(\IMP)}  &\approx  \frac{1}{4 \alpha ^2} \left[\frac{1+e^{2\alpha ^2 \left(1-\eta + \frac{2-\eta }{4 C}\right)}}{2} -\frac{\pi ^2 (1-\eta )}{16 C}\right] \nonumber \\[1ex]
& \hspace{1.cm} -\frac{1}{8 \alpha ^4} \left\{\left[\frac{1+e^{2\alpha ^2 \left(1-\eta + \frac{2-\eta }{4 C}\right)}}{2}\right]^2 - \left[1+\frac{\pi^2}{4}+e^{2\alpha ^2 \left(1-\eta + \frac{2-\eta }{4 C}\right)}\right] \frac{\pi ^2 (1-\eta )}{32 C} \right\} \qquad \mbox{for $\alpha\gg 1$} \,.
\end{align}

%------------%
\subsection{Qubit detection decoherence}
\begin{figure}[t]
\includegraphics[width=0.95\columnwidth]{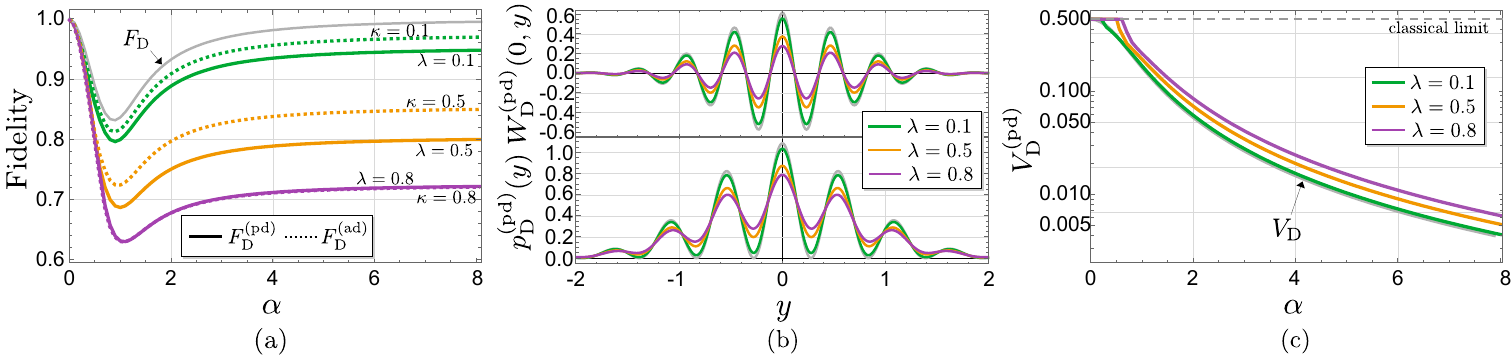}
\caption{(a) Fidelities $F_{\DISP}^{(\rm s)}$, $\rm s= \PD,\AD$, of the dispersive protocol in the presence of qubit decoherence as a function of the target cat amplitude $\alpha$ for different phase damping rate $\lambda\ge 0$ and amplitude damping probability $0\le\kappa\le 1$, respectively. (b) Wigner function $W_\DISP^{(\PD)}(0,y)$ and homodyne probability $p_\DISP^{(\PD)}(y)$ for $\alpha=4$, and (c) distillable variance $V_\DISP^{(\PD)}$ in the presence of phase damping with different $\lambda$. The gray lines refer to the ideal dispersive protocol in the absence of decoherence.}\label{fig:S10-QubitDecoh}
\end{figure}

Being the dispersive protocol also a measurement-based strategy, it is worth to investigate its robustness with respect to decoherence before and inside qubit detection in view of practical applications. To understand the difference to the $\UG(2n)$ scheme mainly limited by PNR detection, we now assume ideal atom-light dispersive coupling, and address the two related noise sources of qubit phase $(\PD)$ and amplitude damping $(\AD)$ of the atom measurement in the $\{|\pm\rangle\}$ basis.

We start by phase damping noise, that models the loss of quantum coherence in the qubit state, and is associated with the quantum completely-positive (CP) map:
\begin{align}
{\cal E}_\PD(\rho)= \frac{1}{\sqrt{4\pi \lambda}} \int_\mathbb{R} d\theta \, e^{-\theta^2/4\lambda} \, e^{-i \theta \sigma_z/2} \, \rho \, e^{i \theta \sigma_z/2} \, ,
\end{align}
$\lambda\ge0$, such that the off diagonal terms of the qubit density matrix are exponentially suppressed by a factor $e^{-\lambda}$, while the diagonal ones are left unchanged \cite{NielsenChuang}. 
Accordingly, in the presence of this damping, the ``atom-light entangled state" $|\Psi_{\DISP} \rangle =(|g\rangle |\alpha\rangle +  |e\rangle |-\alpha\rangle)/\sqrt{2}$ obtained in the ideal dispersive protocol after qubit-light interaction is mapped into the mixture $\Xi_\PD= ({\cal E}_\PD \otimes \Id) (|\Psi_{\DISP} \rangle\langle\Psi_{\DISP}|)$, that subsequently undergoes detection in the $\{|\pm\rangle\}$ basis. The conditional optical states then read:
\begin{align}
\sigma_\pm^{(\PD)} &= \frac{1}{4p_\pm^{(\PD)}(\lambda)} \Bigg\{  |\alpha\rangle \langle \alpha| +  |-\alpha\rangle \langle- \alpha|+ e^{-\lambda} \,\, \Big(|-\alpha\rangle \langle \alpha| + |\alpha\rangle \langle- \alpha| \Big) \Bigg\} \, ,
\end{align}
$p_\pm^{(\PD)}(\lambda)=[1\pm \exp(-\lambda-2\alpha^2)]/2$  being the probability of detecting the qubit in the ``$\pm$" state. 
Thereafter, we perform displacements $D(\pm i\gamma)$ of state $\sigma_-^{(\PD)}$, obtaining the mixture:
\begin{align}\label{eq:rhoPD}
\varrho_\DISP^{(\PD)}(\gamma) = p_+^{(\PD)}(\lambda) \,  \sigma_+^{(\PD)} + \frac{p_-^{(\PD)}(\lambda)}{2} \Big[  D(i\gamma)\, \sigma_-^{(\PD)} D\dag(i\gamma) + D(-i\gamma)\, \sigma_-^{(\PD)} D\dag(-i\gamma) \Big] \, ,
\end{align}
and compute fidelity as:
\begin{align}
F_\DISP^{(\PD)}&= \max_{\gamma} \, \left\langle \cat^{(+)} \right|\varrho_\DISP^{(\PD)}(\gamma) \left|  \cat^{(+)} \right\rangle \, ,
\end{align}
depicted as solid lines in Fig.~\ref{fig:S10-QubitDecoh}(a) for different values of $\lambda$. As expected, decoherence reduces fidelity with respect to the ideal case, but, importantly, $F_\DISP^{(\PD)}$ is still an increasing function of $\alpha$ in the large amplitude regime, and:
\begin{align}
F_\DISP^{(\PD)} \approx \frac{1+e^{-\lambda}}{2} \, \left(1 - \frac{\pi^2}{32 \alpha^2}\right) \qquad \mbox{for $\alpha\gg 1$} \,.
\end{align}
Therefore, in the asymptotic limit, phase damping only appears as a rescaling factor that do not affect the trend of the curve. On the other hand, the Wigner and homodyne distributions experience a reduction of the fringes visibility for larger amounts of damping $\lambda$, see  Fig.~\ref{fig:S10-QubitDecoh}(b): the oscillations of $W_\DISP^{(\PD)}(0,y)$ are progressively suppressed, while the homodyne distribution $p_\DISP^{(\PD)}(y)$ exhibits nonzero minima and reduced heights of maxima, such that for larger $\lambda$ they approach each other and reconstruct a Gaussian envelope in the limit $\lambda\gg1$. In particular, Wigner negativity for large $\alpha$ is reduced as:
\begin{align}
W_\DISP^{(\PD)} (0,\bar{y})\approx W_{\rm cat}^{(+)} (0,\bar{y}) +\frac{2}{\pi} \left(1-e^{-\lambda}\right)+ \frac{\pi}{8\alpha^2} \left(3e^{-\lambda}-2\right) \qquad \mbox{for $\alpha\gg 1$} \,,
\end{align}
with a noise-dependent negativity reduction proportional to $1-e^{-\lambda}$.
Moreover, the distillable squeezing variance $V_\DISP^{(\PD)}$, reported in Fig.~\ref{fig:S10-QubitDecoh}(c), equals the shot noise limit $V_\DISP^{(\PD)}=1/2$ for small amplitudes $\alpha\ll 1$, whilst for larger $\alpha$ it becomes a decreasing function that scales as:
\begin{align}
V_\DISP^{(\PD)} \approx  \frac{1}{4 \alpha ^2}\left(\frac{1+e^\lambda}{2}\right) - \frac{1}{8 \alpha ^4} \left(\frac{1+e^\lambda}{2}\right)^2  \qquad \mbox{for $\alpha\gg 1$} \,.
\end{align}
In the main text, displacement sensitivity of state~(\ref{eq:rhoPD}) is also addressed.

Another detrimental source of noise is provided by amplitude damping, occurring when the atom levels $\{|g\rangle,|e\rangle\}$ in Fig.~\ref{fig:S09-Imperfect}(a) are not degenerate. In such case, spontaneous emission from transition $|e\rangle \to |g\rangle$ is described by the CP map ${\cal E}_\AD(\rho)= \sum_{k=0,1} M_k \, \rho \, M_k\dag$, of Kraus operators $M_0=|g\rangle\langle g| + \sqrt{1-\kappa} |e\rangle\langle e|$ and $M_1= \sqrt{\kappa} |g\rangle \langle e|$, $0\le\kappa\le 1$ being the decay probability \cite{NielsenChuang}. 
In turn, before detection, the output qubit-light state becomes $\Xi_\AD= ({\cal E}_\AD \otimes \Id) (|\Psi_{\DISP} \rangle\langle\Psi_{\DISP}|)$,while the conditional optical state after qubit measurement is equal to:
\begin{align}
\sigma_\pm^{(\AD)} &= \frac{1}{4p_\pm^{(\AD)}(\kappa)}  \Bigg\{  |\alpha\rangle \langle \alpha| +  |-\alpha\rangle \langle- \alpha|+ \sqrt{1-\kappa} \,\, \Big(|-\alpha\rangle \langle \alpha| + |\alpha\rangle \langle- \alpha| \Big) \Bigg\}  \, ,
%\Bigg\{ \sqrt{1-\kappa} \,\, \frac{1\pm e^{-2\alpha^2}}{2} |\cat^{(\pm)}\rangle \langle \cat^{(\pm)}| +\frac{1-  \sqrt{1-\kappa} }{4} \,\, \Big( |\alpha\rangle \langle \alpha| +  |-\alpha\rangle \langle- \alpha| \Big) \Bigg\} \, ,
\end{align}
with $p_\pm^{(\AD)}(\kappa)=[1\pm \sqrt{1-\kappa} \exp(-2\alpha^2)]/2$. After feedforward, the average output state is:
\begin{align}\label{eq:rhoAD}
\varrho_\DISP^{(\AD)}(\gamma) = p_+^{(\AD)}(\kappa) \,  \sigma_+^{(\AD)} + \frac{p_-^{(\AD)}(\kappa)}{2} \Big[  D(i\gamma)\, \sigma_-^{(\AD)} D\dag(i\gamma) + D(-i\gamma)\, \sigma_-^{(\AD)} D\dag(-i\gamma) \Big] \, ,
\end{align}
with optimized fidelity:
\begin{align}
F_\DISP^{(\AD)}&= \max_{\gamma} \, \left\langle \cat^{(+)} \right|\varrho_\DISP^{(\AD)}(\gamma) \left|  \cat^{(+)} \right\rangle \, ,
\end{align}
As shown in Fig.~\ref{fig:S10-QubitDecoh}(a), the effect of amplitude damping is qualitatively equivalent to phase damping, as it still preserves the asymptotic rend of the ideal dispersive scheme, only providing fidelity rescaling in the limit $\alpha\gg 1$, namely: 
\begin{align}
F_\DISP^{(\AD)} \approx \frac{1+\sqrt{1-\kappa}}{2} \, \left(1 - \frac{\pi^2}{32 \alpha^2}\right) \qquad \mbox{for $\alpha\gg 1$} \,.
\end{align}
Analogous considerations can be done for the phase space distributions and distillable squeezing, that exhibit the same behavior depicted in Fig.~\ref{fig:S10-QubitDecoh} (b) and (c) for the phase damping case. In particular, Wigner negativity is suppressed as:
\begin{align}
W_\DISP^{(\AD)} (0,\bar{y})\approx W_{\rm cat}^{(+)} (0,\bar{y}) +\frac{2}{\pi} \left(1-\sqrt{1-\kappa}\right)+ \frac{\pi}{8\alpha^2} \left(3\sqrt{1-\kappa}-2\right) \qquad \mbox{for $\alpha\gg 1$} \,,
\end{align}
while distillable squeezing becomes:
\begin{align}
V_\DISP^{(\PD)} \approx  \frac{1}{4 \alpha ^2} \left(\frac{1+\sqrt{1-\kappa}}{2\sqrt{1-\kappa}}\right) -  \frac{1}{8 \alpha ^4} \left(\frac{1+\sqrt{1-\kappa}}{2\sqrt{1-\kappa}}\right)^2 \qquad \mbox{for $\alpha\gg 1$} \,.
\end{align}

%--------------------------------------------------------------------------------------------------------------------------------------------------%
\section{Quantum Fisher information analysis for displacement sensitivity}\label{sec:FISHER}

\begin{figure}[t]
\includegraphics[width=0.75\columnwidth]{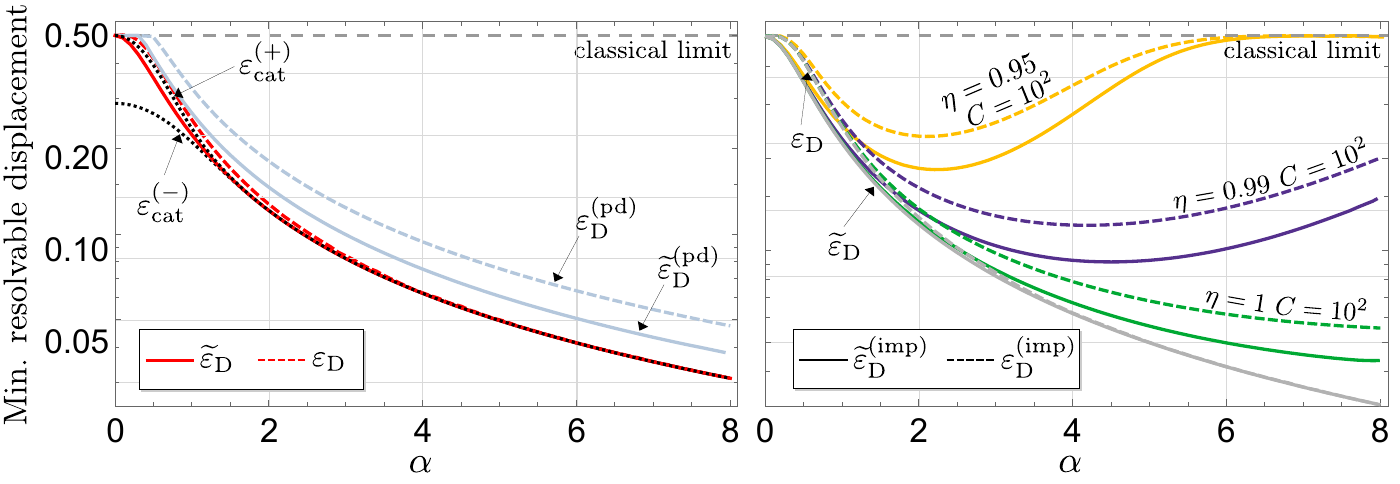}
\caption{Log plot of the minimum resolvable displacements of the ideal (left) and imperfect (right) dispersive couping scheme: solid lines are associated with the quantum Fisher information limit, while dashed lines with momentum-like homodyne detection discussed in the main text. Light colored lines in the left panel refer to the case of qubit phase damping of rate $\lambda=0.2$.}\label{fig:S11-SensingHOMO}
\end{figure}

In the main text, we proposed displacement sensitivity as a figure of merit for the quality of the produced cat states, determining the minimum resolvable displacement $\varepsilon_{\rm min}= 1/\sqrt{\mathbb{F}}$ associated with the classical Fisher information (FI) of momentum-like homodyne detection, computed from the homodyne distributions $p_\DISP(y)$, $p_\DISP^{(\IMP)}(y)$ and $p_\DISP^{(\PD)}(y)$ displayed in Fig.s~\ref{fig:S07-Homodyne},~\ref{fig:S09-Imperfect}(b) and~\ref{fig:S10-QubitDecoh}(b), respectively.

For completeness, here we also present the analysis for the quantum Fisher information (QFI), that provides the ultimate precision limits compatible with quantum mechanics laws for a probe state $\rho_0=\sum_n \rho_n |\varphi_n\rangle\langle \varphi_n|$, $\rho_n\ge 0$, $\langle \varphi_n|\varphi_m\rangle= \delta_{nm}$.
Now, the quantum Cramér-Rao bound determines the optimal minimum resolvable displacement $\widetilde{\varepsilon}_{\rm min}= 1/\sqrt{\mathbb{H}}$, namely the smallest value that the parameter should get to be larger than the quantum-limited minimum uncertainty, where:
\begin{align}\label{eq:QFI}
\mathbb{H}= 4 \, \sum_{nm} \frac{(\rho_n-\rho_m)^2}{\rho_n+\rho_m} \, \left|\langle \varphi_n |\, \hat{x} \, | \varphi_m \rangle \right|^2 \, ,
\end{align}
is the quantum Fisher information (QFI) \cite{Paris2009, Notarnicola2024_Kerr}, being independent of $\epsilon$ as the problem is covariant, and $\hat{x}=(a+a\dag)/\sqrt{2}$ is the position-like operator \cite{NotarnicolaTesi}.

Fig.~\ref{fig:S11-SensingHOMO} reports the values of $\widetilde{\varepsilon}_{\rm min}$ numerically computed for the ideal and imperfect dispersive coupling states $\rho_\DISP$ and $\rho_\DISP^{(\IMP)}$ in Eqs.~(\ref{eq:rhoD}) and~(\ref{eq:rhoIMP}), respectively, as well as for the qubit phase damping noise $\rho_\DISP^{(\PD)}$ in~(\ref{eq:rhoPD}). 
For the target cat states $|\cat^{(\pm)}\rangle$ homodyne detection provides the optimal measurement strategy, with sensitivity
$$\widetilde{\varepsilon}_{\rm cat}^{(\pm)} = \varepsilon_{\rm cat}^{(\pm)} = \frac12 \left( 1+ \frac{4\alpha^2}{1\pm e^{-2\alpha^2}} \right)^{-1/2} \, ,$$
whereas, for the dispersive protocol, homodyning $\hat{y}= i (a-a\dag)/\sqrt{2}$ is only suboptimal, and we have $\widetilde{\varepsilon}_{\rm min} \le \varepsilon_{\rm min}$.
In particular, the ideal coupling reaches intermediate performance between target even and odd cats, $\varepsilon_{\rm cat}^{(-)} \le \widetilde{\varepsilon}_\DISP \le \varepsilon_{\rm cat}^{(+)}$, as state~(\ref{eq:rhoD}) is a convex combination of both even and displaced-odd states, that for $\alpha\gtrsim1.5$ is almost indistinguishable from the target cats. Moreover, in the limit $\alpha \gg 1$, homodyne detection approaches the optimal detection strategy, so that $\widetilde{\varepsilon}_{\DISP} \approx \varepsilon_{\DISP}$.

%--------------------------------------------------------------------------------------------------------------------------------------------------%
\section{Odd cat generation by dispersive protocol}\label{sec:Odd}

\begin{figure}[t]
\includegraphics[width=0.75\columnwidth]{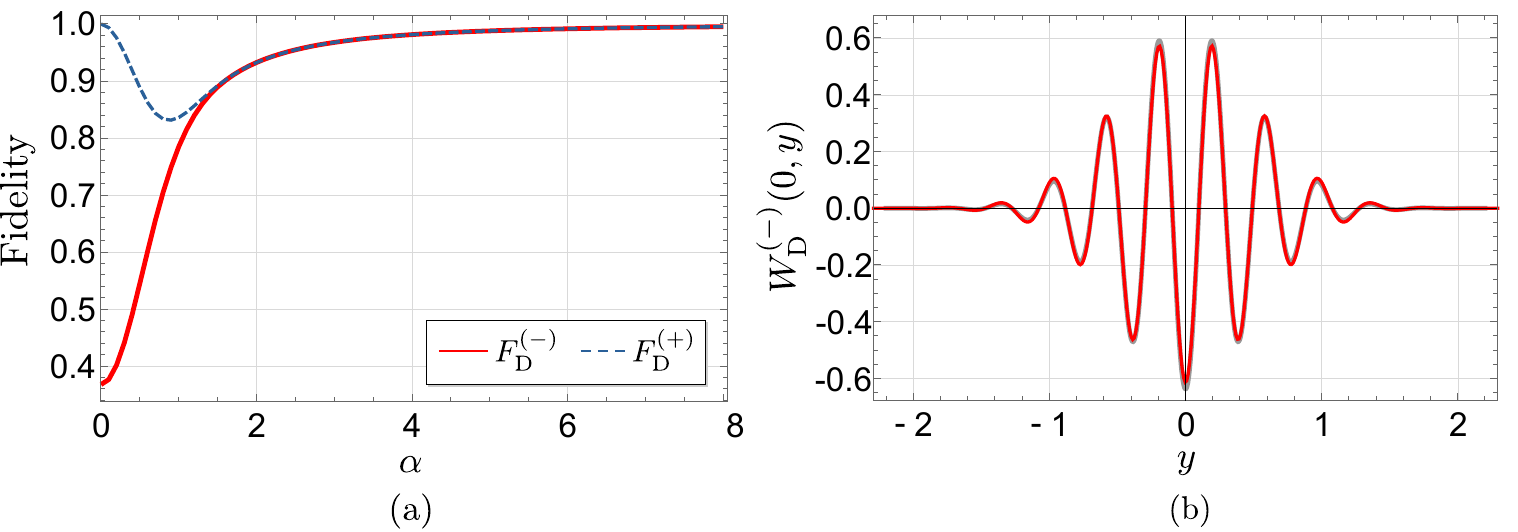}
\caption{(a) Fidelities $F^{(\mp)}_\DISP$ for the odd- and even- cat dispersive protocols as a function of the target cat amplitude $\alpha$. For $\alpha\gg 1$, the odd-cat protocol gets the same scaling $F_\DISP^{(-)} \approx 1 - \pi^2/32 \alpha^2$ as the even cat case discussed in the main text. (b) Oscillations of the Wigner function $W^{(-)}_\DISP(0,y)$ along the $y$ axis for the odd-cat dispersive protocol, in the case $\alpha=4$. The gray line corresponds to the targeted state $|\cat^{(-)}\rangle$. }\label{fig:12-OddCat}
\end{figure}

As a final remark, we note that the dispersive protocol can be straightforwardly extended to also target the odd cat state, namely $|\cat^{(-)}\rangle=(|\alpha\rangle - |-\alpha\rangle)/{\cal N}_-$, with ${\cal N}_-= \sqrt{2(1- \exp(-2\alpha^2))}$. To this aim, we adopt the same protocol depicted in Fig.~1(a) of the main text, where now the feedforward displacement is applied whenever the qubit is found in the $|+\rangle$ state. 
That is, we first exploit dispersive interaction to prepare the ``atom-light entangled state" $|\Psi_{\DISP} \rangle =(|g\rangle |\alpha\rangle +  |e\rangle |-\alpha\rangle)/\sqrt{2}$, followed by qubit measurement in the $|\pm\rangle$ basis. Thereafter, once the qubit is measured in ``-" we create the targeted cat, whilst in the opposite case the orthogonal one is obtained, and we subsequently implement the feedforward displacement $D(\pm i\gamma)$, whose sign is chosen with $50\%$ probability. The output state then reads:
\begin{align}
\varrho_\DISP^{(-)}(\gamma) =  p_- |\cat ^{(-)}\rangle \langle \cat ^{-)}| + p_+ \left( \frac{D(i \gamma)|\cat ^{(+)}\rangle \langle \cat ^{(+)}|D^\dagger(i \gamma) +D(-i \gamma)|\cat ^{(+)}\rangle \langle \cat ^{(+)}|D^\dagger(-i \gamma)}{2} \right) \, ,
\end{align}
where $p_\pm= (1\pm e^{-2\alpha^2})/2$, with corresponding fidelity:
\begin{align}
F_\DISP^{(-)}= \max_{\gamma} \, \left\langle \cat^{(-)} \right|\varrho_\DISP^{(-)} (\gamma) \left|  \cat^{(-)} \right\rangle = p_- +p_+ \left[1+ \coth(2\alpha^2)\right] \max_\gamma \left[ \frac{e^{-\gamma^2}}{2} \sin^2(2\alpha \gamma) \right] \, ,
\end{align}
depicted in Fig.~\ref{fig:12-OddCat}(a) together with the fidelity $F_\DISP^{(+)}$ for the even-cat protocol discussed in the main text. Moreover, oscillations of the corresponding Wigner function $W^{(-)}_\DISP(0,y)$ along the $y$ axis are reported in Fig.~\ref{fig:12-OddCat}(b) in the case $\alpha=4$.

The results are qualitatively similar to the even cat generation. In particular, the optimized displacement amplitude $\gamma_\DISP$ coincides with that of the even-cat protocol, therefore we find the same asymptotic trend 
\begin{align}
F_\DISP^{(-)} \approx F_\DISP^{(+)} \approx 1 - \frac{\pi^2}{32 \alpha^2} \qquad \mbox{for } \alpha\gg 1\,.
\end{align}
Similarly, the Wigner negativity in the origin for large $\alpha$ scales as 
\begin{align}
W_\DISP^{(-)}(0,0) \approx W_{\rm cat}^{(-)}(0,0) + \frac{\pi}{8\alpha^2} \qquad \mbox{for } \alpha\gg 1\,,
\end{align}
with $W_{\rm cat}^{(-)}(0,0)=-2/\pi$. Analogous considerations hold for the distillable squeezing analysis and robustness to imperfect coupling and qubit detection.

%%%%%%%%%%%%%%%%%%%%%%%%%%%%%%%%%%%%%%%%%%%%%%%%%%%%%%%%%
%%%%%%%%%%%%%%%%%%%%%%%%%%%%%%%%%%%%%%%%%%%%%%%%%%%%%%%%%

\bibliography{BiblioCatStates.bib}